\documentclass[]{aa}
\usepackage{graphicx}
\usepackage{subcaption}
\usepackage{txfonts}
\usepackage{amsmath,multirow}
\usepackage{array}
\usepackage{printlen}
\usepackage{caption}
\usepackage{placeins}

\newcommand{\e}{\operatorname{e}}

\newcommand{\bp}{\(\beta\)\,Pic}

\begin{document} 

   \title{
   Search for exocomets transits in Kepler light curves 
   }
   \subtitle{Ten new transits identified}

% \subtitle{I. Overviewing the $\kappa$-mechanism}

   \author{P.~Dumond\inst{1,2}
           \and
           A.~Lecavelier des Etangs\inst{1}
           \and 
           F.~Kiefer\inst{1,3}
           \and
           G.~H\'ebrard\inst{1}
           \and
           V.~Caill\'e\inst{1,4}
           }

    \institute{
    Institut d'astrophysique de Paris, CNRS, Sorbonne Universit\'e, 98$^{\rm bis}$ boulevard Arago, 75014 Paris, FRANCE
    \and
%%    \and
%%
    CRAL, Ecole normale sup\'erieure de Lyon, Universit\'e de Lyon, UMR, CNRS 5574, F-69364 Lyon Cedex 07, France
    \and
    LIRA, Observatoire de Paris, Universit\'e PSL, CNRS, Sorbonne Universit\'e, Universit\'e de Paris, 5 place Jules Janssen, 92195
Meudon, France
    \and
    Institut Pierre-Simon Laplace, Sorbonne-Université, CNRS, Paris, France,
    }
    
%\date{Received September 15, 1996; accepted March 16, 1997}

  \abstract{

The Kepler mission, despite its conclusion over a decade ago, continues to offer a rich dataset for uncovering new astrophysical objects and phenomena. In this study, we conducted a comprehensive search for exocometary transit signatures within the Kepler light curves, 
using a machine learning approach based on a neural network trained on a library of theoretical exocomet transit light curves. 
%derived from {\bf the one published about 25\,years ago.}
%\citet{Lecavelier_1999b}. 
By analyzing the light curves of 201,820 stars, we identified candidate events through the neural network and subjected the output to filtering and visual inspection to mitigate false positives.

Our results are presented into three catalogs of increasing ambiguity. The first-tier catalog includes 17 high-confidence exocometary transit events, comprising 7 previously reported events and 10 newly identified ones, each associated with a different host star. The second-tier catalog lists 30 lower-confidence events that remain consistent with possible exocometary transits. The third-tier catalog consists of 49 more symmetric photometric events that could be either exocometary transits,  exoplanet mono-transits, or false positives due to eclipsing binaries mimicking transits.

Contrary to previous studies, which suggested that the cometary activity was favored by stellar youth, we find a broad age distribution among candidate host stars, including several red giants. This challenges the general idea of a decline in cometary activity with stellar age and underlines the need for further investigation into the temporal evolution of exocometary activity in planetary systems.
}

  \keywords{ Methods: data analysis - Techniques: photometric - Surveys - Comets: general - Planetary systems - Exocomets - Transit photometry }

   \maketitle
%
%-------------------------------------------------------------------
%

\section{Introduction}
\label{Introduction}

Exocomets are rocky or icy minor bodies that become active when they approach their parent star on a highly eccentric orbit \citep{Strom_2020}. Producing a huge tail of gas and dust close to the periastron, when they transit in front of the star, they become detectable using spectroscopy (for the gaseous tail) and photometry (for the dust tail).
The first exocomets were thus identified in the 1980s in the \bp\ system, through the detection of variable absorption features in the Ca\,{\sc ii}\ lines of the stellar spectrum \citep{Ferlet_1987, Kiefer_2014a}. Later on, a few other exocometary systems have been found using spectroscopy, like HD\,172555 \citep{Kiefer_2014b,Grady_2018}, and 49\,Cet \citep{Montgomery_2012,Miles_2016}. A list of spectroscopically identified exocometary systems can be found in \cite{Strom_2020}. 

For the detection of exocomets using photometry, one needs to obtain photometric measurements at high accuracy, typically at $10^{-3}$--$10^{-4}$ level over the full transit duration, that is, for one to three days \citep[][hereafter LdE99a]{Lecavelier_1999a}. This capability has been reached only during the last decade with the space missions like Corot, Kepler, TESS, and Cheops. Detections have thus been obtained with Kepler toward KIC\,3542116  and KIC\,11084727 \citep{Rappaport_2018}, and KIC\,8027456 \citep{Kennedy_2019}, with TESS toward \bp\ \citep{Zieba_2019, Pavlenko_2022, Lecavelier_2022}, and with Cheops toward HD\,172555 \citep{Kiefer_2023}. Here it must also be mentioned the detection with Kepler of a photometric event, possibly periodic, that is interpreted by the transit of a string of 5 to 7~exocomets in front of KIC\,8462852 \citep{Kiefer_2017}.

All these events have been identified thanks to the particular shape of the light curve of an exocomet transit, as theoretically predicted at the end of the 90’s in LdE99a : % by \cite{Lecavelier_1999a}: 
the characteristic of an exocomet transit light curve is the asymmetry caused by the cometary tail passing in front of the star after the nucleus. This causes a sharp decrease of the star light followed by a slow return to the normal brightness \citep[see also][hereafter LdE99b]{Lecavelier_1999b}. However, exocomet transit light curves can also be more symmetric when the longitude of the periastron of the orbit is about 90$^{\circ}$ from the line of sight, because here the cometary tail is aligned with the line of sight. In this case, the shape of the transit light curve is similar to that of an exoplanet transit (LdE99a). %\citep{Lecavelier_1999a}. 

The detection of exocometary transits based on photometry is complementary to the detection through spectroscopy. It allows exploring the physical characteristics of dust particles, such as the dust particle size distribution, the distance to the host star at the time of transit, and the dust production rate \citep{ Lukyanyk_2024}. Indeed, because transits give direct access to the geometrical extent and the optical thickness of the transiting dust cloud, coupled with models of dust production, this allows the estimate of the dust production rate and the size of the comets nucleus. In the case of \bp, statistical analysis of photometric events allowed the estimate of the nucleus size distribution that is found to be similar to that of asteroids and comets in the Solar system as set by collisional equilibrium \citep{Lecavelier_2022}.

Although the characterization of exocomets paves the way to understanding the dynamical and chemical processes occurring in young planetary systems, only a very limited number of exocometary systems have been identified to date.  As a result, our current understanding of the diversity, frequency, and evolutionary significance of exocometary activity remains incomplete. Systematic searches have been performed using spectroscopy in the Ca\,{\sc ii} line \citep{Montgomery_2012, Welsh_2013, Welsh_2018, Rebollido_2020, Bendahan-West_2025}, resulting in a few detections \citep[see review in][]{Strom_2020}. Spectroscopy is efficient in detecting tiny amount of gas and thus allows sensitive search ; the result is that detectable spectroscopic transits seem to be more frequent than detectable photometric events: for instance in the case of \bp, several exocomet transits are routinely detected every day in spectroscopy \citep{Kiefer_2014a}, while only 30~photometric transits have been detected in 156\,days of TESS observations \citep{Lecavelier_2022}. 
However, spectroscopic searches for exocomets are inherently limited by their need to observe each target star individually and the need for long observation time with high spectral resolution. In contrast, photometric surveys can monitor several thousands of stars simultaneously, making them significantly more efficient for large-scale searches for exocometary transits.

In that spirit, \cite{Rappaport_2018} and \cite{Kennedy_2019} have undertaken a deep search for exocometary transits in the Kepler light curves. To identify exocometary transits, they used a modified model of planetary transit or a combination of exponential and Gaussian tails, as also done later on by \cite{Zieba_2019}, \cite{Pavlenko_2022}, and \cite{Lecavelier_2022}. Although this empirical approach does not carry physical information about the cometary bodies and tails, these are simple and easy to implement to identify candidates in the large number ($\sim$200\,000) of light curves provided by Kepler. 

Here we propose to use an automated search without the use of analytical functions to describe the exocometary transit light curves. For that we combined a search algorithm based on a neural network together with a library of photometric transits to train the network. The library is an updated version of the library of LdE99b, %\cite{Lecavelier_1999b}, 
which has been recalculated using faster CPUs than the ones used 20~years ago, covering a wide range of parameters and therefore a wide range of transit shapes that need to be identified. We apply this automated search to the Kepler data \citep{Borucki_2010} with the aim of producing a new catalog of exocometary systems candidates. The methods are described in Sect.~\ref{Training of the neural network}, \ref{The neural network model} and \ref{Evaluation}, the search in Kepler data is presented in Sect.~\ref{Search for exocomets in Kepler data} and the final catalog of candidates, divided in three tiers, is given in Sect.\ref{catalog}. The results are discussed in Sect.\ref{Conclusion}.

\section{Training of the neural network}
\label{Training of the neural network}

\subsection{Training principle}

The method used for the detection of exocomets in Kepler data is based on a neural network using supervised learning. For the network learning, we used a library of theoretical light curves divided into three disjoint sets of labeled light curves: the training set, the validation set, and the test set. Each set contains the same number of light curves with and without an exocomet transit (see Sect.~\ref{light_curve_production}). Each light curve is labeled to indicate if it includes a transit or not.

Our objective is to make the algorithm identify the curves containing an exocomet transit among all the light curves collected by the Kepler satellite and to recover the transit time when a transit is present. Using an iterative procedure, the algorithm optimizes its parameters thanks to the training set. At the end of each iteration step called an "epoch", the network performance is tested using the validation set. This allows us to control the overfitting of the algorithm (e.g. it becomes specific to the training set and loses in generality) and to adjust the hyperparameters. The learning phase is stopped when the network shows the best performance in the validation set. The final performance of the algorithm is given by its evaluation on the test set.

\subsection{Light Curve production}
\label{light_curve_production}
The light curves used to build the training, validation, and test set are the Presearch Data Conditioning (PDC) data from Kepler Data Release 25 \citep{Thompson_2016}. We used the Quarters~1 to~17 and divided them into sub-quarters in case of interruption of the acquisition within a given quarter: two sub-quarters are separated by at least four consecutive photometric measurements missing in the light curve. 

For each sub-quarter, a training, validation, and test set were constructed to overcome their possible specificity. For each sub-quarter, the training set consists of 36\,000 curves, while the validation and test sets contain 12\,000 each. A curve consists on 10 days (480 photometric measurements at a cadence of one measurement every $\sim$30 minutes) of acquisition of the luminosity flux from a star randomly chosen without repetition. 

To ensure efficient learning and prevent the algorithm from falling into obvious traps, several modifications have been made to the Kepler PDC data. First, all points outside the mean value of the light curve by more than 5-sigmas while the neighboring points are not are considered to be outliers and removed. In addition, each curve was normalized to have a zero mean and a standard deviation of 1, as needed for optimal learning by the algorithm. Finally, the possible missing measurements, i.e. at most three consecutive missing measurements within the same sub-quarter, were filled using a linear interpolation. 

For the learning process, half of the 10-day light curves have been used with a label "0", indicating the absence of an exocomet transit: we assume that the probability that any randomly chosen light curve in the Kepler data includes an exocomet transit is extremely low. This set of 0-labeled light curves without exocomet transit is used to learn the noise patterns in the Kepler data. For the other half of the 10-day light curves, we add one exocomet transit by multiplying by a light curve from the theoretical transit library (Sect.~\ref{Simulation of exocomet transits}) and we label them with "1", indicating the presence of an exocomet transit. 

\subsection{Simulation of exocomet transits}
\label{Simulation of exocomet transits}

In order for the algorithm to learn to identify the shape of an exocomet transit in the light curves, we add theoretical exocomet transits to half of the curves that make up the training data. These theoretical exocomet transit light curves have been calculated using the method describes in LdE99a. % \cite{Lecavelier_1999a}. 
We produced a library of 2200 exocomet transit light curves, which is an updated version of the library published twenty years ago by LdE99b. %\cite{Lecavelier_1999b}. 
Among all available simulations, we selected only those with a gas production rate \(\log_{10}(P/(1 \text{kg}\cdot \text{s}^{-1}))\) between 6 and 7.5 and an impact parameter \(b<R_*\), where \(R_*\) is the radius of the star. This choice guarantees that the used transit light curves are sufficiently deep. Indeed, an impact parameter greater than one stellar radius or a too low production rate leads to a noisy and very weak extinction and would therefore have no interest for the training. Before adding the simulated transits to the Kepler light curves, we multiplied the theoretical transit light curve by a random constant such that the depth of the transits is in the interval \([2\sigma, K+2\sigma]\) where \(K\) is the depth of the simulated transits and \(\sigma\) the standard deviation of the Kepler light curves. We thus train our model for a signal-to-noise ratio higher than 2 to exclude the cases where the transit would be non detectable. 
The transit times of the theoretical transits added to the 10-day light curves are chosen randomly, with a uniform distribution between the beginning and the duration of the simulated transit before the end of the light curve.

Finally, in order to ensure that the training data are representative of the diversity of possible transits, we use a uniform random distribution of the distance of the periastron \(q\) in the set \{0.2; 0.3,; 0.5; 0.7; 1.0\} astronomical unit and of the longitude of the periastron \(\omega \) from -157.5$^\circ$ to +157.5$^\circ$ with a step of 22.5$^\circ$,    
as in LdE99b. %\cite{Lecavelier_1999b}. 

The theoretical transits that have been used are totally different for the training, validation and the test sets. To guarantee that the evaluation of the algorithm's performance is unbiased, the choice of the transits for each set has been made randomly from the whole set of theoretical transits.

\section{The neural network model}
\label{The neural network model}

\subsection{Architecture}

The deep learning algorithm developed in this study is mainly based on the convolution layer sequence adapted for pattern recognition. We relied on the architecture of existing algorithms for the detection of exoplanet transits \citep{Shallue_IdentifyingExoplanetsDeep2018, Dattilo_IdentifyingExoplanetsDeep2019,Zucker_ShallowTransitsDeep2018,Chintarungruangchai_DetectingExoplanetTransits2019}. These algorithms, which all have a similar architecture, are optimized for periodic transit detection. They have been modified to be efficient for the detection of single transits, as the orbital period of an exocomet is expected to be much longer than the Kepler mission duration. Two major modifications have been made. "Squeeze-excited blocks" \citep{Hu_SqueezeandExcitationNetworks2018} have been added to the convolution blocks. The authors indeed show that such blocks can improve the performance of a pattern matching algorithm by improving the interdependence relations between the convolution layers without adding computational cost. The addition of the LSTM (Long Short-Term Memory) \citep{Hochreiter_1997_LSTM} and GRU (Gated Recurrent Unit) \citep{cho-etal-2014-learning} layers also resulted in significant performance improvements. 
The algorithm was implemented in TensorFlow, an open-source machine learning framework \citep{Abadi_TensorFlowSystemLargescale2016}. We optimized the neural network parameters thanks to KerasTuner \citep{omalley2019kerastuner}. 
\begin{figure}
   \centering
   \includegraphics[scale=0.8]{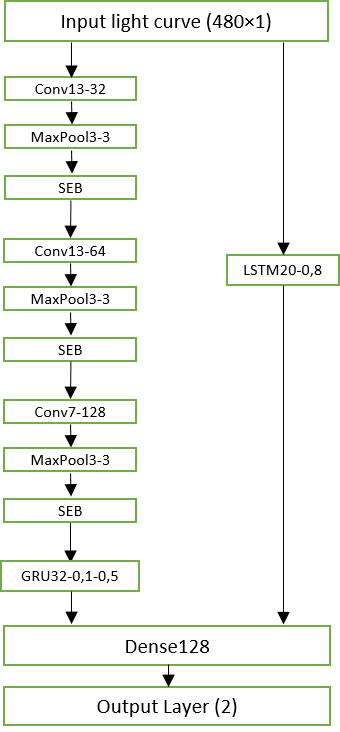}
   \caption{
   Architecture of our best performing neural network model. 
   The input data is a light curve of 480 timesteps, the output is a list of two reals between 0 and 1: one for the probability of detection of an exocometary transit and one for the position of the transit in the light curve.\\
   Convolutional layers are noted {\tt Conv <kernel size> <number of feature maps>}, max pooling layers are noted {\tt MaxPool <window length> <stride length>}, fully connected layers are noted {\tt Dense <number of units>}, LSTM layers are noted {\tt LSTM <units> <dropout>}, GRU layers are noted {\tt GRU <unit> <droupout> <recurent dropout>} and the squeezed-excitation blocks are noted {\tt SEB}.}
   \label{Architecture_neural_net}
\end{figure}

Figure \ref{Architecture_neural_net} describes our final model, which is mainly a one-dimensional convolutional network. The activation function of all the convolutional layers is the {\tt ReLu} (Rectified Linear unit) function \citep{Jarrett_2009_ReLu}. 
The output is a list of two reals. 
The first one uses a sigmoid function whose range is (0,1). It is the probability returned by the algorithm that the 10 days input light curve contains an exocomet transit. An output value close to 1 indicates high confidence that the considered light curve has one transit, while an output value close to 0 means that there is none. The second output layer is activated by a linear function in range (0,1). It gives the position of the identified transit in the light curve. If no transit is detected, the layer should return a value close to zero.

\subsection{Training}
For each sub-quarter, we trained the model using 36\,000~light curves for 150~training epochs, that means that we performed 150~complete pass through the entire training dataset to update the model weight parameters. We used the Adam optimization algorithm \citep{Kingma_AdamMethodStochastic2017} to minimize the cross-entropy error function for the classification of the light curves and the mean absolute error function for the position of the transit. We used a learning rate of \(10^{-4}\) and a batch size of 5\,000. The saved model in the one for which the validation loss reaches its minimum ; usually it is obtained after between 50 and 80 iterations ("epochs"). Therefore, training the model over 150 epochs is considered sufficient. 

\section{Evaluation of the neural network performance}
\label{Evaluation}

We evaluated the performance of our neural network with respect to several metrics in the same way as \cite{Dattilo_IdentifyingExoplanetsDeep2019} did for their model. We computed all metrics over the test set (rather than the training or validation sets) to avoid using any data that were used to optimize the parameters or hyperparameters of the model.
%(see in Sect.~\ref{Metric_histogram}).

\subsection{The metrics}
\label{Metric}
\begin{figure}
   \includegraphics[width=\hsize]{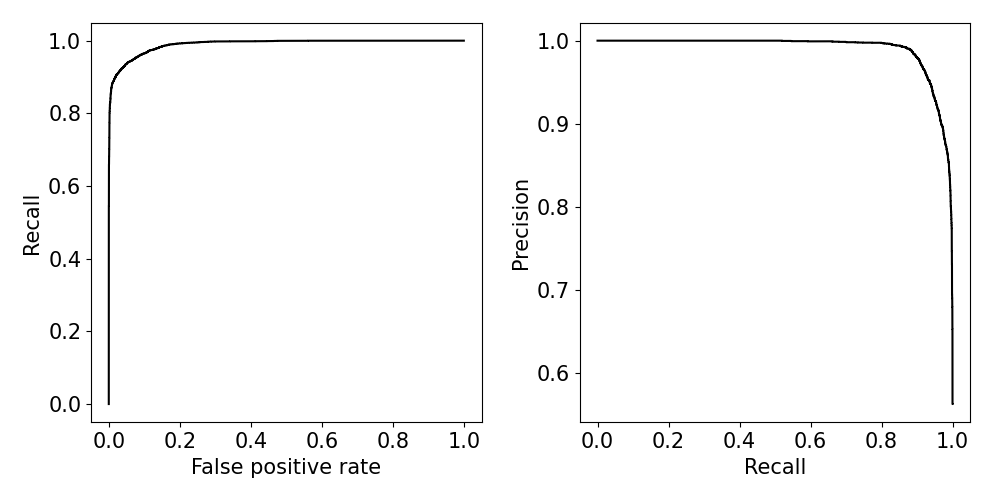}
   \caption[]{
   { The receiver-operator characteristic (ROC) curve (left panel) and the precision versus recall curve (right panel). The receiver-operator characteristic (ROC) curve shows the recall (true-positive rate) of the model against the ability to recognize false positives (the false-positive rate) for different classification thresholds. Our model is highly successful at identifying false positives as shown by the high AUC value  (see Table~\ref{Performances}).
   The plot of the fraction of exocomets that the model classified as exocomets (recall) versus the fraction of correctly classified planets (precision)
   %. This curve 
   shows the trade-off between having no false positives (high precision) and identifying all exocomet transits (high recall).} 
   }
   \label{ROC_curve}
\end{figure}

The metrics used to evaluate the performance of a classifier are the following:
\begin{enumerate}
   \item Accuracy: the fraction of curves correctly classified by the model.
   \item Precision (reliability): the fraction of correctly classified exocomets, i.e., the fraction of curves classified as exocomet candidates (False positives + True positives \(N_{\rm Fp}+N_{\rm Tp}\)) that are indeed true positives  (labeled 1):
   \begin{equation}
      P=\frac{N_{\rm Tp}}{N_{\rm Fp}+N_{\rm Tp}}
   \end{equation}
   \item Recall (completeness or True-positive rate): the fraction of the curves labeled 1, which therefore includes an exocomet transit (True positives + False negatives \(N_{\rm Tp}+N_{\rm Fn}\)), that the model classifies as exocomet candidates ($N_{\rm Tp}$):
   \begin{equation}
      R=\frac{N_{\rm Tp}}{N_{\rm Tp}+N_{\rm Fn}}
   \end{equation}
   \item False-positive rate: the fraction of total labeled 0 curves, which therefore do not include any exocomet transit (False Positives +True Negatives \(N_{\rm Fp}+N_{\rm Tn}\)) that the model classifies as exocomet candidates.
   \begin{equation}
      {\rm FPR}=\frac{N_{\rm Fp}}{N_{\rm Fp}+N_{\rm Tn}}
   \end{equation}
  \item AUC (area under the receiver-operator characteristic curve; see Figure \ref{ROC_curve}): the probability that the model, if it receives two light curves, one 1-labeled curve including an exocomet transit  and one 0-labeled curve without transit, would rank the 1-labeled curve higher than the 0-labeled curve.
  %the probability that a randomly selected exocomet candidate achieves a higher prediction value than a randomly selected false positive. 
    \item Mean Absolute Error: the absolute error for all labeled 1 curves between the true position of the transit and the predicted one by the model.
\end{enumerate}

The first five metrics are related to the detection of the presence of an exocomet in the light curves, while the last metric is related to the position in time of the transit in the light curve. 
Except for the AUC, the values of these metrics depend on the classification threshold chosen for the model. 
This threshold is the value of the probability of detection yield by the algorithm above which we consider that there is a detection.
In Fig.~\ref{ROC_curve}, we show the evolution of the recall as a function of the FPR for various thresholds, usually called the receiver-operator characteristic (ROC) curve. It can be seen that the algorithm differentiates well between true transits and noise, as it achieves very low FPR recall while keeping a reasonably large recall: our model reaches an AUC of 98.88\% for the quarter Q1. 
We also plot the precision as a function of the recall. This curve shows the trade-off between
having no false positives (high precision) and identifying all exocomet transits (high recall).

Considering the huge number of light curves in the Kepler data, the goal is to drastically reduce the number of false positive. We thus chose a classification threshold of 0.99. With this threshold, our model reaches an accuracy of 89.5\% for the quarter Q1. The precision of the model is 99.8\%, which means that we reach a high true positive rate, and the recall is 79,1\%, which means that about 20\% of the true exocomet transits are missing in the final classification. This trade-off aims to accept some loss in the finding of exocomet transits of about 20\% to avoid a large number of false positives. Indeed, given a large number of light curves in the Kepler data and a small number of exocomet transits in these data, even a tiny fraction of false positives can significantly pollute the results. The metrics values are found to be similar for all quarters (see Table~\ref{Performances} in Appendix~\ref{Performance of the algorithm over all the quarters}). 

\subsection{Quality of the output}

\begin{figure}
   \centering
   \includegraphics[width=\hsize]{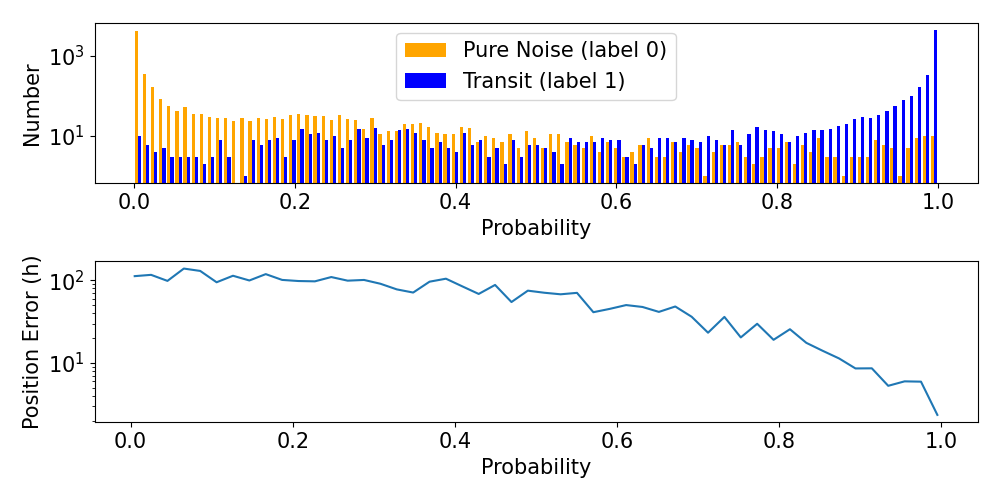}
   \caption{
   {Histogram of the results on the test set and error on the transit position found by the network.}\\
   {\bf Top panel}: Histogram of the result of the neural network applied to the test data. Most of the 1-labeled light curves with an exocomet transit yield a probability close to~1, while there are only a few false positives (0-labeled light curves yielding a high probability). Above the chosen threshold, there is more than two order of magnitude between the number of true positives and false positive. \\
   {\bf Bottom panel}: the mean error of the position of the transit as a function of the probability of the presence of a transit assigned by the network to the light curves. This error is of a few hours at most for the transits identified by the algorithm.}
   \label{Histogram}
\end{figure}

The quality of the model classification can also be visualized by the histogram of the results on the test set (top panel of Fig.~\ref{Histogram}). 
The closer the output from the neural network is to 1 (resp. 0), the higher the probability that the 10-day light curve (resp. does not) contains an exocomet transit. 
The histogram shows that the neural network works as expected: on the left-hand side, most of the light curves characterized by a low probability are mostly noise (orange histogram), while on the right-hand side, the large majority of the curves characterized by a probability close to 1 included a simulated exocomet transit (label 1; blue histogram) and with very few false positives. The bottom graph of Fig.~\ref{Histogram} shows the mean error of the position of the transit as a function of the probability assigned by the algorithm when it is applied to the curves containing a simulated transit (label 1, with known position). One can see that when the transit is not found (probability  close to 0), the error on the predicted position is large, about 3 days, whereas it is of the order of a few hours only when the probability is above 0.99. Thus, one can be confident in the time of the transit given by the algorithm for curves with a probability higher than this threshold.

\section{Search for exocomets in Kepler data}
\label{Search for exocomets in Kepler data}

\subsection{Application of the model to Kepler data}

To apply the algorithm presented in the previous sections, all the light curves of the Kepler data were cut into 10-day windows obtained in the same way as the samples used to construct the training, validation, and test sets described in Sec. \ref{light_curve_production}. In order not to lose any transits located at the edge of the windows, an overlap of 2~days was applied between two consecutive 10-day time intervals.

However, despite the level of precision achieved, the algorithm provides a large number of false positives due to the large number of 10-day windows: approximately $10^6$ windows per quarter lead to about $10^3$ positive candidate detections, among which less than a dozen real transits should be identified (see LdE99a and discussion in Sect.~\ref{Conclusion}). 
To reduce the number of false positives, we decided to filter the candidates using constraints on the shape of the light curves of the detected photometric events.

\subsection{Additional criteria to filter false positives}
\label{Criterions}

After all the Kepler light curves have passed through the neural network, we end up with a set of 56525 exocomet candidates. To filter these candidates, we decided to add constraints on the shape and on the context on the transit (noise, quality flags). 

\subsubsection{Filtering through the shape of the transits candidates}
The shape of the candidate transits were determined by fitting them by the model proposed by \cite{Lecavelier_2022} together with a second order polynomial baseline: 
\begin{equation}
   f_{\rm exo}(t)=\begin{cases} 
      Ct^2+Dt+E & \text{$t < t_0$,} \\
      (Ct^2+Dt+E)(1+ K(\e^{-\beta(t-t_0)}-1)) & \text{$t_1>t > t_0$} \\
      (Ct^2+Dt+E)(1+ K(\e^{-\beta(t-t_0)}-\e^{-\beta(t-t_1)})) & \text{$t > t_1$} 
       \end{cases}
   \label{Model_transit}
\end{equation}
The parameters $t_0$ and $t_1$ represent respectively the time at which the transit starts and the time at which it reaches its minimum. $K$ represents the characteristic depth of the transit, while $1/\beta$ corresponds to the characteristic crossing time of the transit. \(C, D\) and \(E\) are the coefficients of the polynomial baseline. The fit of the light curve by the transit model can be easily performed using the position of the detected candidate in the light curve as provided by the network output.

\begin{figure}
  \resizebox{\hsize}{!}
        {\includegraphics{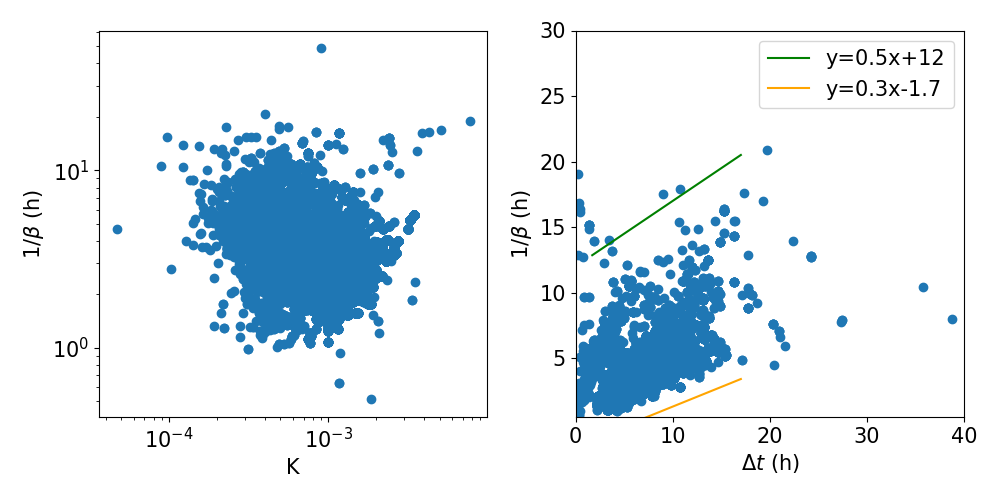}}
  \caption{Distribution of the value of the parameters obtained by fitting the 2163~light curves in the library of the simulated exocomet transits. 
The parameters $K$, \(\Delta T=t_1-t_0\), and $1/\beta$ correspond to the depth, ingress duration, and crossing time of the transit, respectively.
          }
   \label{Coeff_fit}
\end{figure}

To accept or reject a proposed detection given by the neural network, we fit the light curve of the corresponding photometric event to obtain the values of the parameters \(K\), \(\beta\), and \(\Delta T=t_1-t_0\) that characterize the shape of the event. We then check that these values are in the domain of validity for an exocometary transit. The domain of validity is obtained by considering the parameters space occupied by the fits of the theoretical light curves in the library of numerical simulations of exocometary transits, as shown in Fig.~\ref{Coeff_fit} and summarized in the first three lines of Table~\ref{Constrain_fit}. 

It is found that in the theoretical light curves \(K\) is mostly between \(10^{-4}\) and \(5\times 10^{-3}\), \(\Delta T\) is between 1.7 and 17\,hours \citep[corresponding to a periastron between $\sim$0.02 to 2\,au,][]{Lecavelier_2022}, 
and \(\beta^{-1}\) is in the domain above 0.5 hours, and between \(0.3 \Delta T - 1.7\) and \(0.5\Delta T + 12\).
Moreover, the parameters \(C\) and \(D\) have to verify the condition \(|C|, |D|<10^{-3}\). This ensures that the background is stable enough in the neighborhoods of the transit. When this is not the case, the detected transit appears to be only noise. 
% Peut-on donner une indication du nombre ou de la fraction des cas éliminés par ce critère de domaine de validité des paramètres du fit ?

\begin{table*}
   \caption{Parameters values for filtering the candidate transits}             % title of Table
   \label{Constrain_fit}      % is used to refer this table in the text
   \centering                          % used for centering table
   %\begin{tabular}{cm{3.5cm} l l}        % centered columns (3 columns) with vertical alignmen
      \begin{tabular}{c l l l }        % centered columns (3 columns)
   \hline\hline                 % inserts double horizontal lines
   Parameter & Lower bound & Upper bound  & Definition  \\    % table heading 
   \hline \hline                       % inserts single horizontal line
      \(K\) & \(10^{-4}\) & \(5\times 10^{-3}\) & Transit depth \\      % inserting body of the table
      \(\beta^{-1}\) & \(\begin{cases} 0.5\text{h} \\ 0.3\Delta T-1.7 \end{cases}\)  & \( 0.5\Delta T+12 \) & Characteristic crossing time\\
      \(\Delta T\) & 1.7\text{h}     & 17\text{h} & Ingress duration \\
      \(|C|, |D|\) & & \(10^{-3}\) & Baseline coefficients\\
      \(\Delta_{\rm RMS}\) & 36 & \\
      \(\Delta\chi^2_{\rm pl}\) & 5 & \\
      \(\Delta\chi^2_{\rm rp}\) & 10 & \\
   \hline                                   %inserts single line
   \end{tabular}
\end{table*}

\subsubsection{Quality flags}

We noticed that the presence of certain quality flags indicates an alteration of the light curve leading to a systematic detection by the network. Thus, candidates whose transit position is less than 1~day away from a quality flag~3 ("{\it Spacecraft is in coarse point}") or~12 ("{\it Impulsive outlier removed before cotrending}") have been discarded from the candidate list. 

\subsubsection{Presence in the list of KOIs}
To avoid periodic transits, which constitute roughly half of the candidate transits output by the algorithm, stars belonging to the list of KOIs (Kepler Object of Interest) have also been removed.

\subsubsection{Other cases of false positives}
\label{Other cases of false positives}

We have found three other circumstances that can mislead the algorithm and then result in obvious false positives: (1) a significant background noise on a time scale close to that of a cometary transit; (2) a ramp-like shape in the light curve, i.e. a very sudden decrease in flux followed by a more or less rapid return to the normal; (3) the presence of a mono-transit of a possible exoplanet that results in a very symmetrical brightness dip. In the following, we suggest criteria to address each of these three cases.

To characterize the presence of noise anomaly, the \(\Delta_{\rm RMS}\) was calculated. It is defined as:
\begin{equation}
   \Delta_{\rm RMS}=\sum_{i}(f_{\rm exo}(t_i)-\bar{F_3})/\sigma
\end{equation}
where $f_{\rm exo}(t_i)$ is the exocomet model fit to the flux where the candidate transit is found, \(\bar{F_3}\) is the mean of the flux without the three lowest points and \(\sigma\) is the standard deviation of the flux.
Only candidate transits with \(\Delta_{\rm RMS}>36\) were conserved. This criterion was very useful to remove candidates that were detected due to a few points significantly lower than the average. 

To detect the presence of red-noise that mimics an exocomet transit, we calculated the correlation product between the exocomet model fit to the candidate transit and the light curve over all the considered sub-quarters. If the maximum of the correlation \(\text{max}_1\) is reached more than one day away from the transit time or if its second maximum \(\text{max}_2\) is significantly close to the first maximum (\((\text{max}_1-\text{max}_2)/\text{max}_1\)\,<\,0.15), the candidate is removed, considering that in this case the noise in the light curve resembles the identified transit, which is therefore probably not a real exocomet transit, but a background red-noise. 

The cases of the ramp-like light curves mimicking an exocometary transit have been discarded by calculating the difference between the fit with a comet model (Eq.~\ref{Model_transit}) and a fit with a simple ramp model defined by: 
\begin{equation}
   f_{\rm ramp}(t)=\begin{cases} 
      Ct^2+Dt+E & \text{$t < t_0$,} \\
      (Ct^2+Dt+E)(1- K\e^{-\beta(t-t_0)}) & \text{$t > t_0$} 
       \end{cases}
   \label{Model_ramp}
\end{equation}
We calculate \(\Delta\chi^2_{\rm rp}\) defined by :
\begin{equation}
   \Delta\chi^2_{\rm rp}=N\left(\frac{\sum_{i}(f_{\rm ramp}(t_i)-F(t_i))^2}{\sum_{i}(f_{\rm exo}(t_i)-F(t_i))^2}-1\right),
\end{equation}
where \(N=120\) is the number of points in the fitted window. 

The photometric events that are almost as well fitted with the ramp model as with the exocomet model (\(\Delta\chi^2_{\rm rp}<10\)), have been discarded from the candidate list. 

Finally, to identify the cases of a symmetrical transit that could be due to the passage of an exoplanet with a long period (a monotransit), 
we calculate the difference \(\Delta\chi^2_{\rm pl}\) between the \(\chi^2_{\rm pl}\) of a model of a planetary transit calculated using the equations of  \cite{Mandel_AnalyticLightCurves2002} and the \(\chi^2_{\rm exo}\) of an exocomet transit model (Eq.~\ref{Model_transit}). It can be written as:
\begin{equation}
   \Delta\chi^2_{\rm pl}=N\left(\frac{\sum_{i}(f_{\rm pl}(t_i)-F(t_i))^2}{\sum_{i}(f_{\rm exo}(t_i)-F(t_i))^2}-1\right).
\end{equation}
The photometric events having \(\Delta\chi^2_{\rm pl}<5\) have a symmetrical shape and have therefore not been taken into account in our final selection of exocomets. Such transits have been taken into account to establish the list of symmetrical mono-transits (see Sec. \ref{Sec_monotransits}). This criterion is, however, not used as a sharp cut to distinguish the exocomets transits from the symmetrical mono-transits but only as a help. The final distinction has been made through visual inspection. 

After applying these criteria, the number of candidate transits per quarter varies between 50 and 100 per quarter, leading to a total of 1349 transits. It then becomes possible to proceed for each of them to a visual inspection to know if the candidate is indeed or not an exocomet transit. This final inspection is necessary because, although the criteria put in place make it possible to eliminate the vast majority of false positives, there still remain obvious false positives. Note here that the criteria have been deliberately chosen to be conservative so as not to accidentally eliminate a true exocomet transit. The number of criteria was chosen so that the final number of transits to be visually inspected would be no more than a few thousand. Otherwise, visual inspection would be impractical. It would be more inefficient and inaccurate to add extra criteria than to visually check the current list of 1,349 transits.

{To verify the robustness of the detected transits, we perform an additional test. We apply our detection procedure to the data after reversing the time. If the proportion of detected exocomets is the same in the reversed time data as in the original data, this may question the relevance of the detected transits.
However, as this analysis is time-consuming, we performed this analysis on half of the quarters. Of the 17 quarters, we analysed eight randomly chosen quarters: Q1, Q2, Q6, Q9, Q10, Q12, Q15 and Q16. We found only one transit that clearly mimics an exocomet transit: KIC\_3129239 in Q9. Given that approximately 17 robust transits have been detected in the Kepler data (see Sec. \ref{catalog}), we can infer from this study that about 10\% to 15\% of these transits can be false positives. 
Although the same number of detections in the time-reversed data would have disproven the exocomet hypothesis, the result of this test supports the idea that there is an astrophysical signal in the data that is consistent with exocomet transits. 
}

\section{A new catalog of transiting exocomets}
\label{catalog}

In this section, we present the final list of exocomet transits resulting from the selection procedure. We divided the list in three different tiers based on the likelihood of a genuine exocometary detection. In the first tier, the photometric events are the ones that we consider as the most likely due to the transit of an exocomet~; in the second tier are the events possibly due to comets, while the third tier gathers the most symmetric dip in the light curve, for which the transit of an exoplanet remains a possibility. 

\subsection{The first-tier catalog of exocometary transits}

The first-tier catalog of detected exocometary transits is given in Table~\ref{Detections} and the plots of the corresponding light curves are given in Appendix~\ref{Cometary transits light curves}. It is particularly noteworthy that our neural network is able to find all the transits already identified by previous works: KIC\,3542116 in Quarters~1, 10 and~12 and KIC\,11084727 as found by \cite{Rappaport_2018}, and KIC\,8027456 as found by \cite{Kennedy_2019}.  
This independent retrieval of the same exocometary transits provides confidence in our procedure. 

Nonetheless, after application of all the search procedure described above, it appeared that two transits of KIC\,3542116 identified by \cite{Rappaport_2018} in Quarters~8 were not in our list of detections. 
These transits were eliminated by the criterion based on the maximum of the correlation product that aims to remove the detections whose pattern occurs several times in the same subquarter (Sect.~\ref{Other cases of false positives}). 
In fact, this criterion also has the disadvantage of eliminating active cometary systems where at least two comets transits occurred in the same quarter, as is the case for KIC\,3542166 in Quarter~8.
Therefore, we made a visual inspection of all photometric events eliminated because of this criterion in order to recover possible obvious exocometary transits that were wrongly rejected.
We thus found the two transits of KIC\,3542116 in the Quarter~8, and in addition identify a new transit in front of KIC\,6263848. This last case does not correspond to a star with frequent cometary transits, but had been deleted due to an edge effect on the calculation of the correlation function.
These two cases are identified by an asterisk in Table~\ref{Detections}. 

For all of these detection, we calculate the $\Delta\chi^2_{\text{polynomial}}$ defined as
\begin{equation}
   \Delta\chi^2_{\rm polynomial}=N\left(\frac{\sum_{i}(f_{\rm polynomial}(t_i)-F(t_i))^2}{\sum_{i}(f_{\rm exo}(t_i)-F(t_i))^2}-1\right).
\end{equation}
where $f_{\rm polynomial}$ is the second order polynomial that best fits the detection. For all detections, we obtained large $\Delta\chi^2_{\text{polynomial}}$ (larger than about 100). This highlights that the detection cannot be interpreted just as noise patterns.

The properties of the parent stars of the identified exocomets
as tabulated by Kepler are given in Table~\ref{tab:kepler_stars} 
\citep{Brown_2011}. For KIC\,3542116 and KIC\,3662483, $T_{\rm eff}$, \(\log_{10} g\), metallicity and radius are from \cite{Zhang_2025}. 
From the \(T_{\rm eff}\) and \(\log _{10} g\) we can obtain an estimate of the stellar type \citep{Gray_2008}, which is given in the last column of Table~\ref{tab:kepler_stars}.
It appears that most of the stars are main-sequence stars, with the exception of KIC\,4078638, which is likely a red giant. 
For these stars, the upper limits on the age estimate given by \cite{Zhang_2025} are all above 10$^9$\,years, with the exception of KIC\,4481029 with an upper limit of \(9\times 10^{8}\)\,years, and KIC\,8027456 with an age estimate of \(5\times 10^{8}\)\,years.
%
%\\
%catalog name following LSS 
%\\

\begin{table*}
   \centering
   \caption{First-tier catalog of exocometary transits. }
   \label{Detections}
   \begin{tabular}{r r r r r r r r r }
   \hline
   \hline
       Stars (KIC) & Time (BKJD) & Quarter & K$(\times 10^{-4})$ & $\beta^{-1} (h)$ & $\Delta T (h)$ & AD$(\times 10^{-4})$  & \(\Delta\chi^2_{\text{pl}}\) & \(\Delta\chi^2_{\text{polynomial}}\) \\ 
\hline 
$^{\dag}$3542116 & 161.5 &1 &4.74 &6.5 &9.92 &3.71 &18.32 &279.74 \\ 
$^{\dag}$*3542116 & 742.6 & 8 & 4.17 &1.36 & 6.05 & 4.12 & 0.6 & 348.83 \\ 
$^{\dag}$*3542116 & 792.9 & 8 & 8.62 & 4.16 & 4.77 & 5.88 & 15.57 & 285.27 \\ 
$^{\dag}$3542116 & 992.1 & 10 & 10.12 & 3.62 & 14.68 & 9.94 & 28.67 & 2993.05 \\ 
$^{\dag}$3542116 & 1175.8 & 12 & 12.98 & 3.40 & 9.35 & 12.15 & 8.58 & 2670.38 \\ 
3662483 & 1132.1 & 12 & 6.81 & 6.38 & 6.87 & 4.49 & 12.98 & 204.52 \\ 
4078638 & 355.5 & 4 & 83.21 & 6.30 & 3.22 & 33.30 & 47.89 & 358.90 \\ 
4481029 & 172.5 & 2 & 9.08 & 8.57 & 4.64 & 3.80 & 14.31 & 235.74 \\ 
5206257 & 925.0 & 10 & 10.98 & 4.75 & 4.47 & 6.69 & 5.04 & 322.26 \\ 
5514200 & 670.1 & 7 & 34.48 & 12.13 & 7.27 & 15.53 & 19.15 & 155.20 \\ 
*6263848 & 775.9 & 8 & 10.22 & 3.29 & 4.90 & 7.92 & 7.96 & 151.93 \\ 
6927963 & 416.8 & 4 & 33.60 & 9.13 & 2.83 & 8.96 & 6.54 & 115.89 \\ 
7660548 & 966.2 & 10 & 15.74 & 5.95 & 5.09 & 9.05 & 26.10 & 219.47 \\ 
$^{\dag}$8027456 & 1449.0 & 15 & 8.0 & 9.90 & 14.12 & 6.08 & 16.78 & 401.18 \\ 
8738545 & 238.1 & 2 & 17.0 & 4.92 & 3.67 & 8.94 & 7.64 & 92.53 \\ 
10484683 & 309.5 & 3 & 29.85 & 14.05 & 8.01 & 12.97 & 87.66 & 853.82 \\ 
$^{\dag}$11084727 & 1076.2 & 11 & 15.29 & 3.92 & 8.20 & 13.40 & 88.14 & 2368.80 \\ 
\hline
\hline
   \end{tabular}
   \tablefoot{The first column gives the name of the star by its KIC identifier. The second and third columns indicate the position of the transit. Columns 4 to 6 give the fits parameters associated with the exocomet model (Eq.~\ref{Model_transit}). The seventh column gives the absorption depth. The last two columns present the difference of \(\chi^2\) between the cometary model and, first, the planetary model \citep{Mandel_AnalyticLightCurves2002} and, second, a second order polynomial fit.\\
   The transits that were first eliminated by the criterion based on the maximum of the correlation product (see Sect.~\ref{Other cases of false positives}) are identified by an asterisk (*). 
   The five transits already known are identified by the $^{\dag}$ symbol.}
   \end{table*}

\begin{table*}[h]
    \centering
    \caption{Information on the stars of the first-tier catalog.}
    \label{tab:kepler_stars}
    \begin{tabular}{r c c c c c c c c}
\hline
\hline

\textbf{Kepler ID} & \textbf{RA} & \textbf{Dec} & \textbf{Kepler Mag} & \textbf{Teff} & \textbf{Log g} & \textbf{Metallicity} & \textbf{Radius} & \textbf{Stellar} \\
\textbf{} & \textbf{(J2000)} & \textbf{(J2000)} & \textbf{} & \textbf{(K)} & \textbf{(cm/s$^2$)} & \textbf{} & \textbf{(R$_\odot$)} & \textbf{type} \\
\hline
3542116       & 19 22 52.939   & +38 41 41.51     & 9.979          & 6766   & 4.208  & -0.190  & 1.454  & F2 V                 \\
3662483       & 19 42 09.199   & +38 43 01.38     & 10.541         & 5848   & 3.945  & 0.107   & 1.786  & G1 IV                 \\
4078638       & 19 46 16.611   & +39 06 15.23     & 13.156         & 4783   & 2.758  & 0.150   & 8.331  & K3 III            \\
4481029       & 19 43 13.210   & +39 32 32.03     & 11.303         & 8865   & 3.968  & 0.010   & 2.310  & A0 IV             \\
5206257       & 19 46 33.826   & +40 22 57.97     & 12.051         & 6017   & 4.435  & -0.105  & 1.052  & F6 V              \\
5514200       & 18 58 08.306   & +40 46 29.32     & 14.377         & 5651   & 4.500  & -0.243  & 0.955  & G4 V              \\
6263848       & 18 52 30.566   & +41 36 33.34     & 13.172         & 5850   & 4.365  & -0.224  & 1.137  & G2 V              \\
6927963       & 18 56 15.946   & +42 27 52.20     & 13.807         & 5711   & 4.441  & -0.012  & 1.030  & G4 V              \\
7660548       & 18 47 22.807   & +43 23 05.39     & 13.094         & 6158   & 4.369  & -0.050  & 1.148  & F5 V              \\
8027456       & 19 25 15.838   & +43 51 33.55     & 9.697          & 8732   & 3.766  & 0.065   & 2.974  & A0 IV             \\
8738545       & 18 58 37.262   & +44 55 15.24     & 13.227         & 5091   & 3.581  & -0.162  & 3.101  & K1 IV            \\
10484683      & 19 47 54.377   & +47 40 14.16     & 12.588         & 6326   & 4.543  & -0.009  & 0.934  & F5 V              \\
11084727      & 19 28 41.191   & +48 41 15.14     & 9.987          & 6762   & 4.067  & -0.152  & 1.726  & F3 V           \\
\hline
\hline   
   \end{tabular}
\end{table*}

\subsection{The second-tier catalog of possible exocometary}

After visual inspection of the list of candidates produced by our algorithm, we came up with a list of interesting photometric events that can be due to transiting exocomets. This includes transits that pass through all the criteria described in Sect.~\ref{Search for exocomets in Kepler data}, but whose shape or signal-to-noise ratio makes it difficult to decide whether they are actually transits. The resulting list of possible exocometary transits in this second-tier catalog is given in Table~\ref{Possible_detections}  and the plots of the corresponding light curves are given in Appendix~\ref{Possible exocometary transits light curves}.

The properties of the corresponding parent stars in the second tier catalog are given in Table~\ref{tab:kepler_star_tier_2} 
\citep{Brown_2011}. 
Here again, estimates of the stellar types given in the last column of the table show that most of the stars are on or close to the main sequence, with the exception of KIC\,2984102 and KIC\,11153134, which have a \(\log _{10} g\) below~3 and are therefore likely red giants. 
For all stars in this second-tier catalog, the upper limit of the age estimate given by \cite{Zhang_2025} is above 10$^9$\,years, with the exception of KIC\,2984102, KIC\,5294231 and KIC\,7183123 with age estimates of \(5\times 10^{8}\), \(1\times 10^{8}\) (with an upper limit of \(3\times 10^{8}\)) and \(9\times 10^{8}\)\,years, respectively.

\begin{table*}
   \centering
   \caption{Second-tier catalog of possible exocometary transits. Same legend as in Table~\ref{Detections}.}
   \label{Possible_detections}
   \begin{tabular}{r r r r r r r r r}
   \hline
   \hline
       Stars (KIC) & Time (BKJD) & Quarter & K$(\times 10^{-4})$ & $\beta^{-1} ({\rm h})$ & $\Delta T ({\rm h})$ & AD$(\times 10^{-4})$  &  \(\Delta\chi^2_{\text{pl}}\) & \(\Delta\chi^2_{\text{polynomial}}\) \\ \hline \hline
       2984102 & 345.7 & 3 & 36.42 & 15.9 & 10.12 & 17.14 & 61.90 & 430.26 \\ 
4826941&385.8&4&21.58&15.22&7.85&8.69&3.76&126.17 \\ 
5195476 & 1508.8 & 16 & 16.80 & 5.09 & 3.67 & 8.31 & 3.42 & 45.92 \\ 
5294231 & 411.4 & 4 & 4.92 & 3.26 & 2.13 & 2.36 & 31.78 & 285.46 \\ 
5511746 & 1105.8 & 12 & 13.31 & 12.33 & 5.06 & 4.48 & 2.41 & 44.54 \\ 
5514277 & 670.4 & 7 & 63.11 & 13.57 & 9.88 & 32.64 & 13.96 & 160.95 \\ 
5534083 & 1167.7 & 12 & 9.34 & 10.45 & 3.33 & 2.55 & 7.22 & 43.16 \\ 
6119914 & 747.2 & 8 & 31.41 & 9.36 & 6.63 & 15.95 & 3.54 & 50.65 \\ 
7183123 & 336.4 & 3 & 8.43 & 8.36 & 5.98 & 4.31 & 28.32 & 321.42 \\ 
7201827 & 409.9 & 4 & 43.64 & 4.83 & 2.83 & 19.34 & 5.77 & 48.98 \\ 
7220968 & 1325.7 & 14 & 18.60 & 5.50 & 6.23 & 12.61 & 2.61 & 46.42 \\ 
7611858 & 1275.8 & 14 & 6.46 & 9.77 & 5.19 & 2.67 & 7.63 & 109.47 \\ 
7669613 & 315.9 & 3 & 13.24 & 13.66 & 6.31 & 4.89 & 12.77 & 121.72 \\ 
8111649 & 934.0 & 10 & 83.04 & 23.69 & 10.28 & 29.24 & 16.36 & 123.61 \\ 
8144412 & 639.1 & 7 & 86.39 & 12.00 & 2.95 & 18.84 & 7.06 & 140.49 \\ 
8374877 & 769.1 & 8 & 7.78 & 7.04 & 8.46 & 5.44 & 13.67 & 300.35 \\ 
8489495 & 1351.8 & 14 & 16.43 & 8.44 & 15.04 & 13.67 & 10.70 & 47.92 \\ 
8625941 & 1505.0 & 16 & 5.16 & 8.44 & 9.80 & 3.54 & 8.10 & 66.25 \\ 
8683719 & 1340.2 & 14 & 19.21 & 6.43 & 14.25 & 17.12 & 7.59 & 195.79 \\ 
8738569 & 1462.3 & 15 & 13.18 & 8.30 & 4.30 & 5.33 & 8.84 & 56.11 \\ 
8843356 & 702.2 & 7 & 8.84 & 10.54 & 2.99 & 2.18 & 4.66 & 46.51 \\ 
9025688 & 1078.2 & 11 & 13.29 & 7.51 & 3.57 & 5.03 & 58.03 & 58.03 \\ 
9388975 & 680.3 & 7 & 18.41 & 3.36 & 2.47 & 9.59 & 6.76 & 60.28 \\ 
9531080 & 748.0 & 8 & 10.23 & 9.31 & 6.56 & 5.18 & 16.13 & 53.73 \\ 
10329957 & 1385.1 & 15 & 18.47 & 12.90 & 4.10 & 5.03 & 9.62 & 64.04 \\ 
11124157 & 704.2 & 7 & 15.01 & 10.93 & 2.00 & 2.51 & 11.48 & 45.20 \\ 
11153134 & 739.7 & 8 & 18.93 & 3.15 & 7.99 & 17.44 & 11.42 & 345.83 \\ 
11246607 & 696.4 & 7 & 18.06 & 15.10 & 8.84 & 8.00 & 7.04 & 165.29 \\ 
11515196 & 644.3 & 7 & 16.40 & 7.98 & 3.84 & 6.26 & 34.15 & 138.30 \\ 
11701976 & 931.7 & 10 & 23.79 & 11.98 & 2.70 & 4.80 & 13.24 & 40.42 \\ 
       
   \end{tabular}
\end{table*}

\begin{table*}[h]
    \centering
    \caption{Information on the stars of the second-tier catalog.}
    \label{tab:kepler_star_tier_2}
    \begin{tabular}{r c c c c c c c c}
    \hline \hline
\textbf{Kepler ID} & \textbf{RA} & \textbf{Dec} & \textbf{Kepler Mag} & \textbf{Teff} & \textbf{Log g} & \textbf{Metallicity} & \textbf{Radius} & \textbf{Stellar} \\
\textbf{} & \textbf{(J2000)} & \textbf{(J2000)} & \textbf{} & \textbf{(K)} & \textbf{(cm/s$^2$)} & \textbf{} & \textbf{(R$_\odot$)} & \textbf{type} \\
    \hline \hline
          2984102& 19 20 24.302  & +38 09 41.26  &11.271&5064&2.543&-0.167& 10.715 & K1III \\
          4826941& 19 16 42.927  & +39 58 17.69  &13.917&5748&4.425&-0.257& 1.051	& G3V\\
          5195476& 19 36 47.573  & +40 19 41.48  &13.925&4848&3.504&-0.214& 3.406 & K2IV \\
          5294231& 19 46 54.941  & +40 26 00.17  &9.842&9552&4.082&0.032& 2.107 & A0V \\
          5511746& 18 53 09.113  & +40 46 26.87  &13.216&5928&4.266&-0.277& 1.290 & G1V \\
          5514277& 18 58 16.380  & +40 47 18.02  &15.366&5401&4.601&-0.242& 0.824 & G7V \\
          5534083& 19 27 57.571  & +40 45 16.38  &12.602&6416&4.186&-0.261& 1.458 & F4V \\
          6119914& 19 22 29.443  & +41 24 30.31  &15.882&6167&4.819&-0.036& 0.660 & F5V \\
          7183123& 19 02 20.712  & +42 47 30.55  &11.207&8005&3.703&-0.145& 3.020 & A0IV \\
          7201827& 19 29 03.487  & +42 44 03.41  &15.409&5562&4.815&-0.098& 0.638 & G5V \\
          7220968& 19 48 59.026  & +42 44 39.52  &15.463&5015&4.483&0.35& 0.921 & K1V \\
          7611858& 19 32 36.857  & +43 17 11.54  &11.187&5684&4.477&-0.315& 0.984 & G4V \\
          7669613& 19 06 56.856  & +43 18 44.17  &13.044&5754&4.246&-0.382& 1.313 & G3V \\
          8111649& 19 45 47.110  & +43 58 36.44  &15.776&6080&4.576&-0.044& 0.888 & F6V \\
          8144412& 18 47 41.122  & +44 04 04.40  &15.213&5721&4.428&0.043& 1.047 & G3V \\
          8374877& 19 40 27.934  & +44 19 09.62  &12.083&6510&4.021&-0.505& 1.809 & F4V \\
          8489495& 19 18 40.234  & +44 33 44.75  &15.214&4912&4.328&0.068& 1.112 & K2V \\
          8625941& 19 28 55.855  & +44 44 53.77  &12.783&5689&4.602&-0.318& 0.844 & G4V \\
          8683719& 19 18 58.464  & +44 49 28.02  &14.717&5989&4.398&-0.035& 1.100 & G0V \\
          8738569& 18 58 41.198  & +44 57 51.12  &12.771&4907&3.548&0.173& 3.227 & K2IV \\
          8843356& 19 58 40.882  & +45 01 02.64  &12.039&5419&3.694&-0.387& 2.687 & G7IV \\
          9025688& 19 32 42.634  & +45 23 41.57  &13.382&6445&4.165&-1.598& 1.500 & F4V \\
          9388975& 18 56 00.802  & +45 57 05.90  &14.292&5163&4.493&-0.938& 0.923 & K0V \\
          9531080& 19 35 07.706  & +46 09 35.64  &13.861&6280&4.268&-0.346& 1.306 & F5V \\
          10329957& 19 10 18.223  & +47 27 04.43  &13.58&5002&4.567&-0.024& 0.824 & K1V \\
          11124157& 18 56 39.605  & +48 45 44.21  &12.244&4775&3.954&0.071& 1.869 & K3IV \\
          11153134& 19 53 11.854  & +48 46 20.82  &12.007&4941&2.719&-0.037& 8.746 & K2III \\
          11246607& 19 32 21.149  & +48 58 47.78  &13.169&6251&4.387&-0.253& 1.127 & F5V \\
          11515196& 19 42 06.725  & +49 26 57.88  &13.568&4754&4.15&0.345& 1.403 & K3V \\
          11701976& 19 02 02.592  & +49 49 17.65  &13.403&6210&4.398&-0.014& 1.111 & F5V \\
    \hline \hline
    \end{tabular}
\end{table*}

\subsection{The third-tier catalog of symmetric transits}
\label{Sec_monotransits}

After visual inspection of the list of candidates produced by our algorithm, we also came up with a list of interesting photometric events that appears to be symmetrical. This is not surprising as some of the exocomets transits in the library used for training the network are indeed symmetrical (see Fig.~2 and~3 in LdE99a). %\citep[see Fig.~2 and~3 in][]{Lecavelier_1999a}. 
However, we cannot exclude that these are due to mono-transits of exoplanets or to false positives mimicking planetary or cometary transits. Several scenarios might cause such false positives, including diluted or undiluted eclipsing binaries. To identify these cases one needs to carry out additional analyses that are beyond the scope of the present paper \citep[see, e.g.,][]{Crossfield_2016}. 
Finally, although an exocometary transit classification cannot be made, these detections deserve to be mentioned. They are given in the third-tier catalog of symmetric transits listed in Table~\ref{Table_Symetric_transits}.

We compared these detected photometric events with the lists of single transits published by \cite{Huang_2013}, \cite{Wang_2015}, \cite{Foreman-Mackey_2016}, and \cite{Herman_2019}. It appears that four events in our list were already identified by \cite{Foreman-Mackey_2016} and \cite{Herman_2019}. The transit in front of KIC\,8410697 was found in both studies. It is interpreted as due to an exoplanet with a radius of 0.7~times that of Jupiter. The photometric event of KIC\,10321319 was found by \cite{Foreman-Mackey_2016} and interpreted as due to the transit of an exoplanet of 0.16 Jupiter radius. The photometric event of KIC\,6196417 was found by \cite{Herman_2019} and interpreted as due to the transit of an exoplanet of about 0.7 Jupiter radius. 
The transit in front of KIC\,10668646 at the time of ${\rm BKJD}$=1449.3 was already identified by \cite{Foreman-Mackey_2016} but the exoplanet transit scenario was rejected because of a centroid shift in the data. However, we also identified another photometric event in the same target at ${\rm BKJD}$=196.3 with a different shape and a lower absorption depth. 
These few examples show that our list of newly identified photometric events which could be due to transits of exocomet with symmetrical shape of the light curve deserves further investigations which are beyond the scope of the present paper. 

\begin{table*}
   \centering
   \caption{Third-tier catalog of symmetric transits that could be either exocometary transits or exoplanet mono-transits.}
   \label{Table_Symetric_transits}
   \begin{tabular}{r r r r r r r r r}
   \hline
   \hline
Stars (KIC) & Time (BKJD) & Quarter & K$(\times 10^{-4})$ & $\beta^{-1} ({\rm h})$ & $\Delta T ({\rm h})$ & AD$(\times 10^{-4})$  &  \(\Delta\chi^2_{\text{pl}}\) & \(\Delta\chi^2_{\text{polynomial}}\) \\ \hline \hline
2983000&759.1&8&18.38&0.84&16.33&18.38&-1.94&483.82 \\ 
2993038&550.9&6&15.22&2.46&4.72&12.99&-1.57&1165.8 \\ 
3222471&673.1&7&97.7&5.49&14.59&90.83&-70.29&2436.25 \\ 
3346436&270.4&3&8.48&3.03&3.05&5.38&3.42&182.43 \\ 
3755854&1492.3&16&3.53&1.07&6.11&3.51&1.03&45.73 \\ 
5184479&534.6&5&8.87&1.04&19.61&8.87&-11.23&316.53 \\ 
5184479&1203.0&13&12.17&1.97&17.91&12.17&-3.69&409.71 \\ 
5305217&673.1&7&33.0&5.46&20.87&32.28&-7.41&170.59 \\ 
5456365&270.8&3&35.42&3.85&10.37&33.03&-2.28&323.73 \\ 
5905878&1343.3&14&2.25&1.46&7.55&2.24&-7.35&353.12 \\ 
5905878&1441.3&15&0.89&1.49&8.14&0.89&-2.01&60.16 \\ 
5967153&1394.5&15&5.06&4.05&7.9&4.34&-22.09&141.91 \\ 
5975275&213.3&2&9.78&2.57&10.57&9.62&2.7&179.4 \\ 
$^{\tt b}$6186417&959.1&10&33.19&1.09&18.42&33.19&-2.36&645.89 \\ 
6387193&651.7&7&78.79&2.73&12.43&77.96&-48.64&3817.28 \\ 
6515488&1420.4&15&28.85&2.98&9.54&27.67&-7.02&359.44 \\ 
6804821&1009.1&11&12.97&1.6&22.41&12.97&-6.09&317.96 \\ 
7047396&298.8&3&11.35&2.0&5.57&10.65&-4.56&305.5 \\ 
7105703&1509.0&16&20.54&2.86&6.53&18.45&-4.32&186.11 \\ 
7213651&1239.2&13&21.53&1.66&6.78&21.17&-9.23&214.24 \\ 
7465971&342.7&3&11.14&1.14&9.75&11.14&-1.07&200.89 \\ 
7983622&471.9&5&65.06&4.63&13.76&61.73&-15.71&732.08 \\ 
8005892&608.5&6&96.11&1.2&9.78&96.09&-75.32&9905.87 \\ 
8007462&1458.6&15&2.44&2.56&5.4&2.14&-7.15&93.52 \\ 
8159297&440.7&4&18.1&1.87&4.62&16.57&-1.87&132.24 \\ 
8313257&1149.1&12&25.05&2.17&17.4&25.05&-8.72&320.72 \\ 
$^{\tt a,b}$8410697&542.5&6&54.16&1.0&17.87&54.16&-12.07&5694.83 \\ 
8496108&277.7&3&19.41&9.57&3.83&6.4&-6.55&62.5 \\ 
8617888&171.8&2&55.2&1.83&5.44&52.4&-4.99&149.3 \\ 
9016734&666.0&7&61.61&5.57&13.7&56.35&-48.25&1624.94 \\ 
9016734&1230.1&13&27.52&5.5&14.31&25.48&-7.51&522.79 \\ 
9413755&1564.2&17&19.5&1.63&4.4&18.18&-16.75&217.06 \\ 
9775416&1398.1&15&5.04&4.99&10.63&4.44&-2.36&141.64 \\ 
10024862&1494.0&16&20.95&1.68&13.29&20.94&-2.57&121.87 \\ 
10157075&369.5&4&26.75&1.93&5.57&25.27&-2.17&114.74 \\ 
10192453&757.5&8&18.64&5.63&1.73&4.94&-29.2&76.45 \\ 
$^{\tt a}$10321319&554.5&6&4.76&4.77&13.56&4.48&-29.18&174.87 \\ 
10334763&549.7&6&7.23&0.74&10.88&7.23&-3.39&412.26 \\ 
10397849&506.5&5&19.93&2.35&7.95&19.25&0.15&289.02 \\ 
10450889&239.6&2&60.03&1.58&2.35&46.44&1.45&192.69 \\ 
10556420&756.3&8&3.59&0.64&6.03&3.59&-1.17&364.07 \\ 
10643786&1421.2&15&12.62&3.23&5.12&10.04&3.16&81.51 \\ 
10668646&196.3&2&42.81&0.93&8.74&42.8&-20.57&1376.58 \\ 
$^{\tt a}$10668646&1449.3&15&57.08&1.36&8.64&56.98&-38.4&3632.55 \\ 
10909733&898.7&9&3.41&1.37&6.17&3.37&4.64&67.65 \\ 
11561379&137.7&1&13.51&2.11&7.45&13.12&-1.64&103.03 \\ 
11921843&1234.8&13&3.85&0.62&14.83&3.85&-2.49&287.33 \\ 
12066509&632.3&7&38.71&0.78&12.41&38.71&-0.31&1883.85 \\ 
12644038&584.7&6&36.26&1.49&6.37&35.77&3.56&199.23 \\ 
\hline\hline
   \end{tabular}
\tablefoot{Four transits were already identified by \cite{Foreman-Mackey_2016} (labeled with "{\tt a}" in the first column) and \cite{Herman_2019}
   (labeled with "{\tt b}").}
\end{table*}

\section{Conclusion}
\label{Conclusion}

Although the Kepler mission ended more than 10~years ago, the available data is still worth deep data mining in search of new discoveries. Here we present a new search for exocometary transits in the Kepler light curves. We used machine learning technique with a neural network that has been trained using a library of theoretical exocomets transit light curves inherited from the work of LdE99b. %\cite{Lecavelier_1999b}.  
After parsing the light curves of close to 200\,000~stars through the neural network, despite the application of several filters to eliminate most of the false positive, a visual inspection of the outcome of the network was still needed. We ended up with three catalogs of interesting objects. The first-tier catalog is composed of a total of 17~exocometary transits, including 7~previously identified transits and 10~new transits in front of 10~different stars. The second-tier catalog is a list of 30~photometric events that appear of second quality and that we qualify as possible exocometary transits. Finally, the third-tier catalog presents the list of interesting photometric events that are symmetrical and may be due to transits of either an exoplanet or an exocomet with a periastron at 90$^\circ$ from the line of sight.

The complete Kepler data represent a 4-year photometry survey of more than 170\,000 distinct stars. However, as noted in \cite{Kennedy_2019},
not all stars were observed for the full duration of the mission. 
As a result, we can consider 150\,000 as the approximate number of equivalent stars that were observed for the full duration of the mission. 
Or, in other words, our analysis covers the equivalent of about 600\,000\,stars-years.
%\cite{Lecavelier_1999a} showed that 
LdE99a showed that 
with 30\,000\,stars-years survey of stars with a solar system cometary activity and a few $10^{-4}$ photometric accuracy, the number of exocomets transit detections should be around 10. Therefore, assuming that each star harbors a cometary system similar to our own, we could expect roughly 200~detections with a perfect search in the Kepler data. Finding a few dozens of exocometary transits can therefore be considered as a satisfactory result in agreement with what could be expected using reasonable estimates of the algorithm efficiency and cometary activity of the surveyed stars. 

Unfortunately, the Kepler stars are rather faint, and our catalogs contain stars with magnitude between 9.7 and 15.9. It is therefore not possible to undertake a spectroscopic follow up with the hope to confirm the exocometary nature of the photometric events. All the more that, with only one transit per star in about four years, exocometary transits are rare and not predictable in time.
However, we can anticipate that the same techniques could be applied to search for exocometary transits in current TESS data and in upcoming PLATO data. For these two instruments, the surveyed star in the input catalog are significantly brighter that the Kepler stars. We can thus expect that the exocometary systems to be discovered in near future can be studied in detail, particularly through spectroscopy.

Previous searches \citep{Rappaport_2018,Kennedy_2019} provided a list of 3~stars that seem to be rather young. This supported the anticipated conclusion that cometary activity is likely correlated with stellar age, with an expected decline over time. However, our catalogs do not support this idea, as we found a wide spread of age for both the first- and second-tier catalogs, with possible exocometary transits in front of stars classified as red giants. The issue of cometary activity with age thus remains open, and further studies will be needed to clarify this important question of the evolution of planetary systems. 

\textbf{Note added in manuscript}: {After this manuscript was submitted we became aware of a paper by \cite{Norazman_2025} that used Machine Learning techniques to search exocometary transits in the sectors 1 to 22 of the TESS light curves. They found three additional transits compared to those already known, which underlines the utility of this technique for the search of such transits.}

\begin{acknowledgements}

We thank Dr.~Neda Heidari for her help in identifying already known mono-transits.  
We acknowledge support from the
CNES (Centre national d’études spatiales, France).
This work has made use of the Infinity Cluster hosted by Institut d'Astrophysique de Paris. We thank Dr.~Guilhem Lavaux for his invaluable help in making efficient use of the GPUs in this cluster.
This paper includes data collected by the Kepler mission and obtained from the MAST data archive at the Space Telescope Science Institute (STScI). Funding for the Kepler mission is provided by the NASA Science Mission Directorate. STScI is operated by the Association of Universities for Research in Astronomy, Inc., under NASA contract NAS 5–26555.
This research has made use of the NASA Exoplanet Archive, which is operated by the California Institute of Technology, under contract with the National Aeronautics and Space Administration under the Exoplanet Exploration Program.

\end{acknowledgements}

\bibliographystyle{aa}
\bibliography{Bib_exo}

\newpage
\onecolumn
\begin{appendix} %First appendix
   \section{Performance of the algorithm over all the quarters}
   \label{Performance of the algorithm over all the quarters}
\begin{table}[h!]
    \centering
      \caption{Performance of the neural network.}
      \label{Performances}
    \begin{tabular}{l l l l l l l l }
      \hline\hline
          Quarter & Subquarter & Accuracy & Recall & Precision & FPR & AUC & MAE \\ 
        \hline \hline
          1 & 0 &  0.8945  &  0.7910  &  0.9975  &  0.0020  &  0.9888  &  2.3404  \\ \hline
          2 & 0 &  0.8665  &  0.7347  &  0.9977  &  0.0017  &  0.9907  &  2.2237  \\ \hline
          2 & 1 &  0.8616  &  0.7245  &  0.9982  &  0.0013  &  0.9900  &  2.0792  \\ \hline
          2 & 2 &  0.8743  &  0.7498  &  0.9984  &  0.0012  &  0.9893  &  2.0455  \\ \hline
          3 & 0 &  0.9108  &  0.8240  &  0.9972  &  0.0023  &  0.9896  &  2.0329  \\ \hline
          3 & 1 &  0.9068  &  0.8162  &  0.9969  &  0.0025  &  0.9905  &  2.1186  \\ \hline
          3 & 2 &  0.8903  &  0.7835  &  0.9964  &  0.0028  &  0.9895  &  2.1638  \\ \hline
          4 & 0 &  0.8862  &  0.7745  &  0.9972  &  0.0022  &  0.9909  &  2.3073  \\ \hline
          4 & 1 &  0.9017  &  0.8065  &  0.9961  &  0.0032  &  0.9924  &  2.4661  \\ \hline
          4 & 2 &  0.8854  &  0.7728  &  0.9974  &  0.0020  &  0.9920  &  1.9370  \\ \hline
          5 & 0 &  0.8994  &  0.8013  &  0.9969  &  0.0025  &  0.9919  &  1.9726  \\ \hline
          5 & 1 &  0.8898  &  0.7818  &  0.9972  &  0.0022  &  0.9916  &  2.1857  \\ \hline
          5 & 2 &  0.8951  &  0.7923  &  0.9973  &  0.0022  &  0.9919  &  2.5873  \\ \hline
          6 & 0 &  0.8979  &  0.7972  &  0.9983  &  0.0013  &  0.9906  &  2.2020  \\ \hline
          6 & 1 &  0.8908  &  0.7833  &  0.9979  &  0.0017  &  0.9911  &  1.9720  \\ \hline
          6 & 2 &  0.8860  &  0.7738  &  0.9976  &  0.0018  &  0.9910  &  2.4570  \\ \hline
          7 & 0 &  0.8792  &  0.7605  &  0.9972  &  0.0022  &  0.9899  &  2.1862  \\ \hline
          7 & 1 &  0.8752  &  0.7522  &  0.9978  &  0.0017  &  0.9909  &  1.7915  \\ \hline
          7 & 2 &  0.9128  &  0.8267  &  0.9986  &  0.0012  &  0.9935  &  2.1101  \\ \hline
          8 & 0 &  0.9004  &  0.8018  &  0.9988  &  0.0010  &  0.9911  &  2.0729  \\ \hline
          8 & 1 &  0.9027  &  0.8065  &  0.9986  &  0.0012  &  0.9926  &  2.1139  \\ \hline
          9 & 0 &  0.9068  &  0.8157  &  0.9976  &  0.0020  &  0.9930  &  2.1293  \\ \hline
          9 & 1 &  0.9017  &  0.8062  &  0.9965  &  0.0028  &  0.9916  &  2.3157  \\ \hline
          9 & 2 &  0.8983  &  0.7992  &  0.9969  &  0.0025  &  0.9918  &  2.3348  \\ \hline
          9 & 3 &  0.8798  &  0.7610  &  0.9980  &  0.0015  &  0.9916  &  2.2389  \\ \hline
          10 & 0 &  0.8755  &  0.7532  &  0.9971  &  0.0022  &  0.9896  &  2.7444  \\ \hline
          10 & 1 &  0.8758  &  0.7538  &  0.9971  &  0.0022  &  0.9889  &  2.0156  \\ \hline
          10 & 2 &  0.8829  &  0.7672  &  0.9983  &  0.0013  &  0.9890  &  2.7258  \\ \hline
          11 & 0 &  0.8830  &  0.7685  &  0.9968  &  0.0025  &  0.9908  &  2.0722  \\ \hline
          11 & 1 &  0.8941  &  0.7900  &  0.9977  &  0.0018  &  0.9913  &  1.9576  \\ \hline
          11 & 2 &  0.8907  &  0.7828  &  0.9981  &  0.0015  &  0.9921  &  2.4971  \\ \hline
          12 & 0 &  0.8876  &  0.7767  &  0.9981  &  0.0015  &  0.9921  &  2.3743  \\ \hline
          12 & 1 &  0.9028  &  0.8075  &  0.9977  &  0.0018  &  0.9930  &  2.3531  \\ \hline
          12 & 2 &  0.9038  &  0.8103  &  0.9967  &  0.0027  &  0.9913  &  2.4029  \\ \hline
          13 & 0 &  0.8981  &  0.8000  &  0.9952  &  0.0038  &  0.9913  &  2.8353  \\ \hline
          13 & 1 &  0.8431  &  0.6875  &  0.9981  &  0.0013  &  0.9872  &  2.4796  \\ \hline
          13 & 2 &  0.9037  &  0.8117  &  0.9947  &  0.0043  &  0.9915  &  2.5127  \\ \hline
          14 & 0 &  0.8654  &  0.7342  &  0.9955  &  0.0033  &  0.9877  &  2.4772  \\ \hline
          14 & 1 &  0.8801  &  0.7632  &  0.9961  &  0.0030  &  0.9897  &  2.4096  \\ \hline
          14 & 2 &  0.8788  &  0.7592  &  0.9978  &  0.0017  &  0.9915  &  3.1279  \\ \hline
          15 & 0 &  0.8363  &  0.6743  &  0.9973  &  0.0018  &  0.9869  &  2.5724  \\ \hline
          15 & 1 &  0.8378  &  0.6780  &  0.9966  &  0.0023  &  0.9826  &  3.3750  \\ \hline
          15 & 2 &  0.8490  &  0.7000  &  0.9972  &  0.0020  &  0.9862  &  3.1479  \\ \hline
          16 & 0 &  0.8238  &  0.6483  &  0.9987  &  0.0008  &  0.9870  &  2.4968  \\ \hline
          16 & 1 &  0.8665  &  0.7355  &  0.9966  &  0.0025  &  0.9868  &  2.1849  \\ \hline
          16 & 2 &  0.8445  &  0.6907  &  0.9976  &  0.0017  &  0.9858  &  2.4707  \\ \hline
          17 & 0 &  0.8744  &  0.7510  &  0.9971  &  0.0022  &  0.9899  &  2.3075 \\ \hline
        \end{tabular}
    \end{table}
   
\newpage
    \section{Cometary transits light curves}
    \label{Cometary transits light curves}
Plots of the light curves of the detected cometary transits in the first tier catalog. For each transit, the light curve is shown in the top panels over a 5~days duration centered on the transit time and with an exocomet transit fit superimposed (red line). The bottom panels show the light curves over a larger time frame. 
\begin{figure*}[h!]
\begin{subfigure}{.5\hsize}
    \centering
    \includegraphics[width=\hsize]{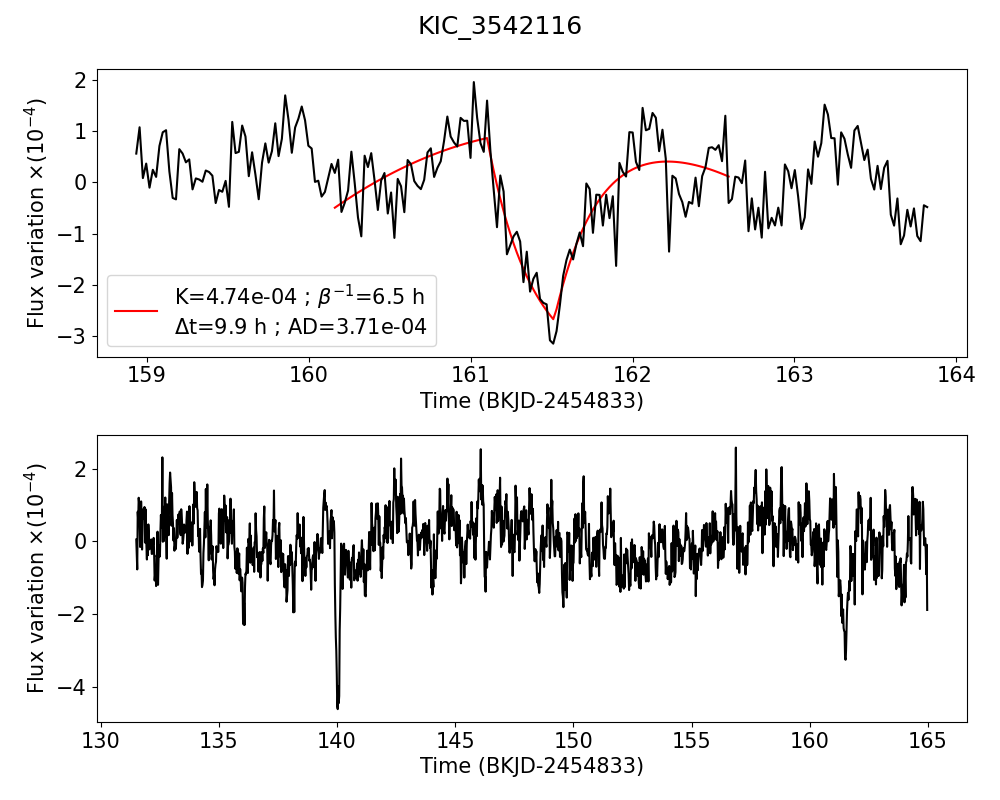}
\end{subfigure}
\begin{subfigure}{.5\hsize}
    \centering
    \includegraphics[width=\hsize]{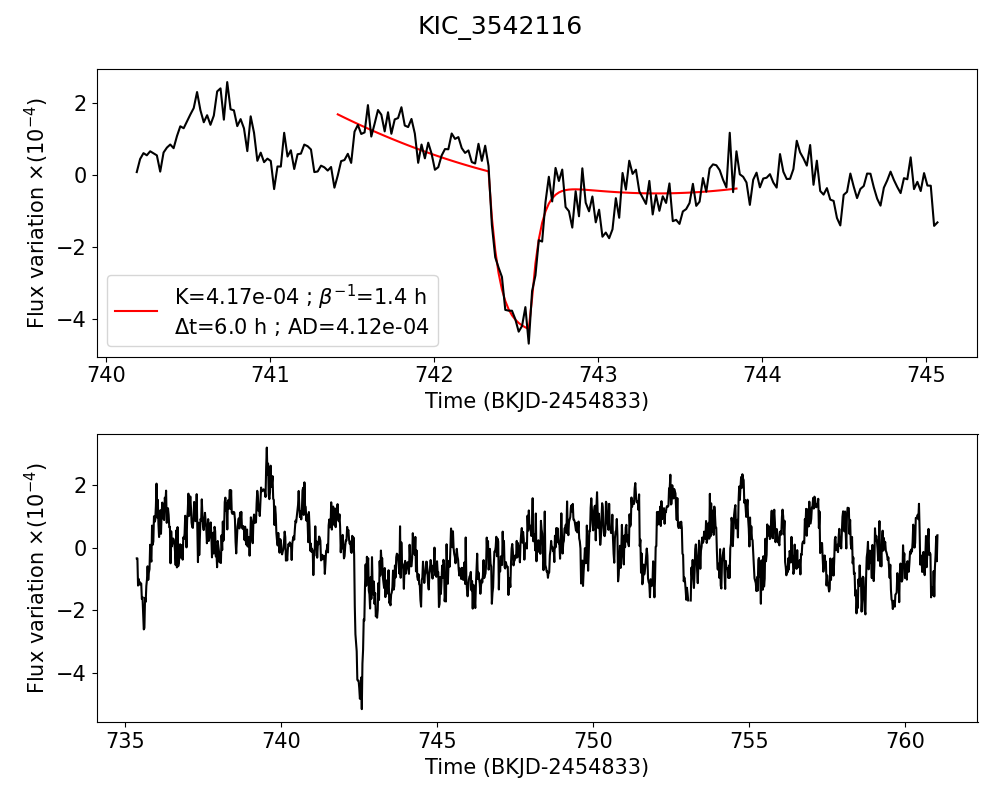}
\end{subfigure}
\begin{subfigure}{.5\hsize}
    \centering
    \includegraphics[width=\hsize]{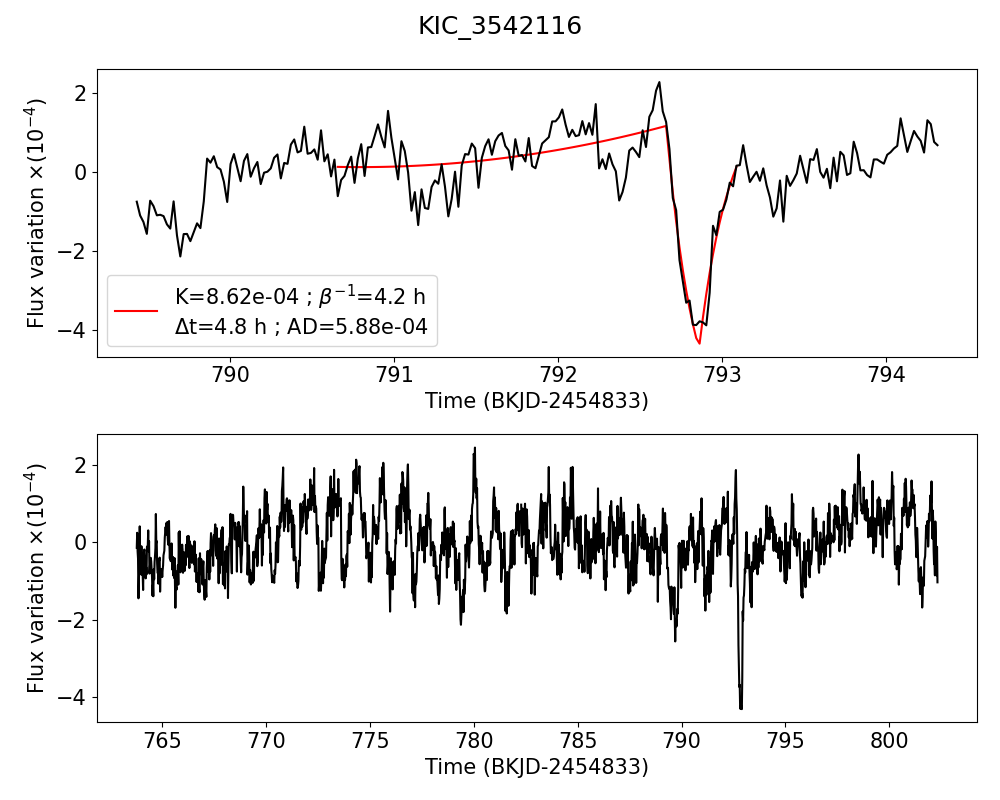}
\end{subfigure}
\begin{subfigure}{.5\hsize}
    \centering
    \includegraphics[width=\hsize]{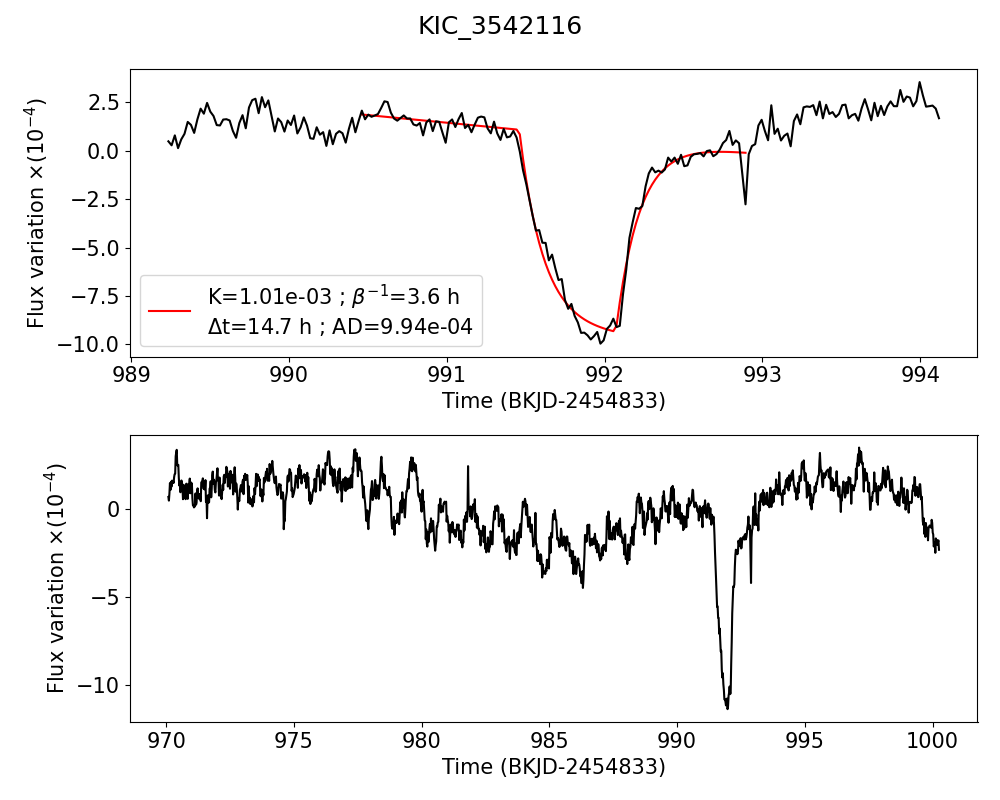}
\end{subfigure}
\begin{subfigure}{.5\hsize}
    \centering
    \includegraphics[width=\hsize]{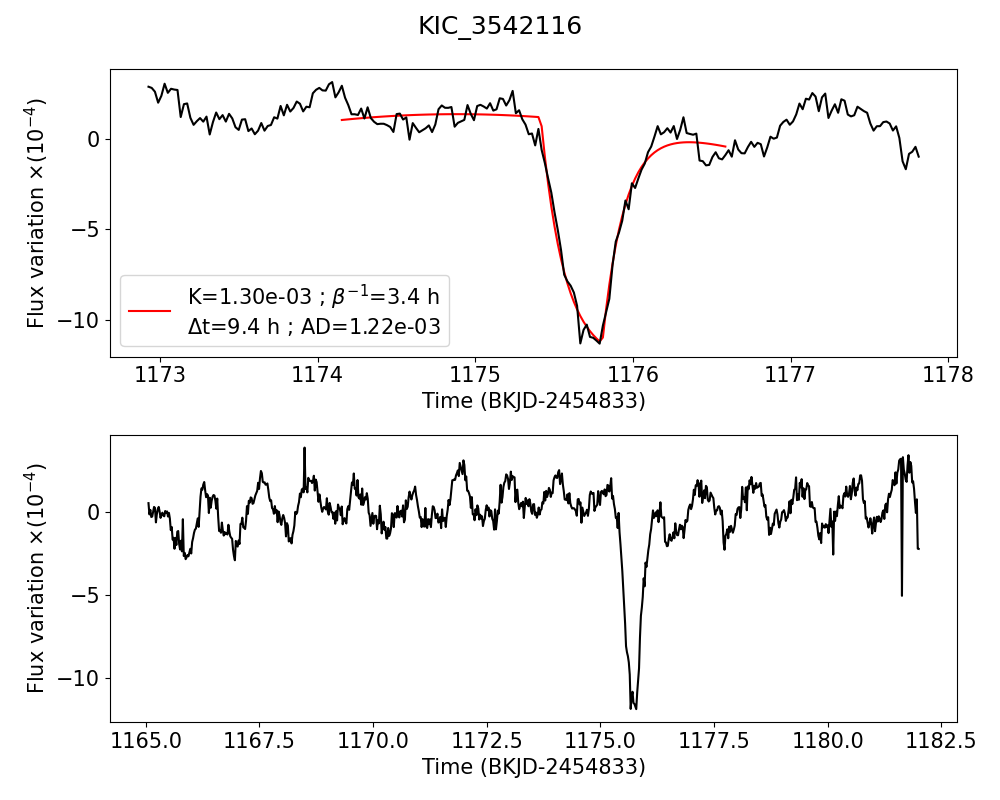}
\end{subfigure}
\begin{subfigure}{.5\hsize}
    \centering
    \includegraphics[width=\hsize]{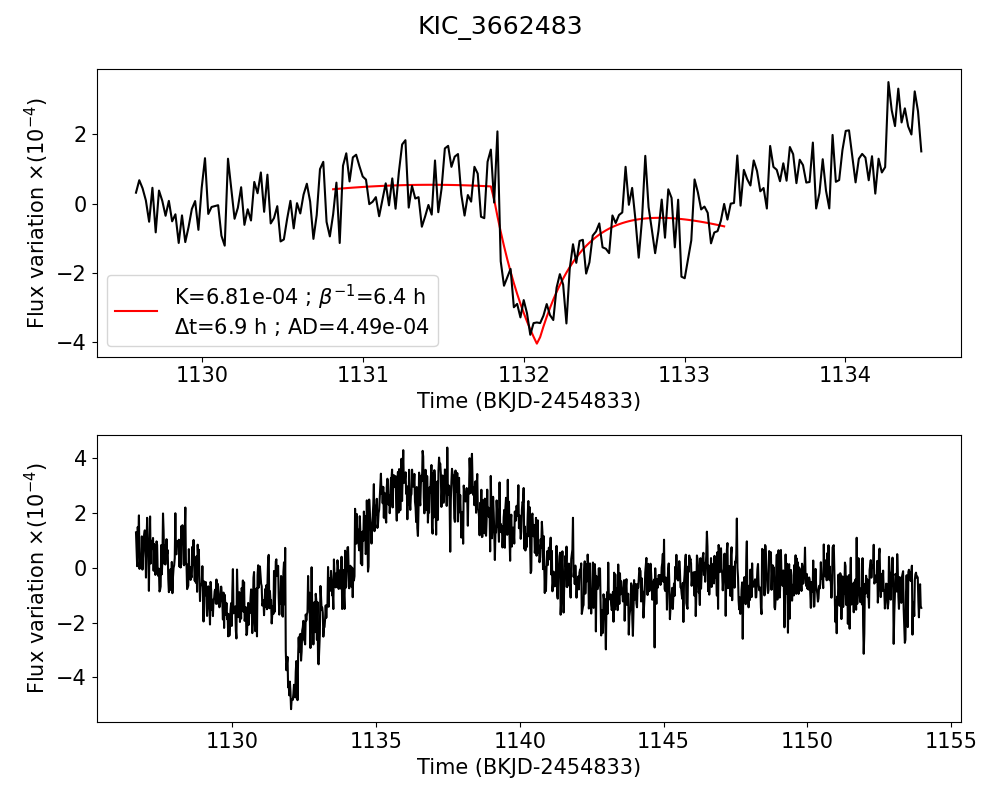}
\end{subfigure}
\end{figure*}

\begin{figure*}
\begin{subfigure}{.5\hsize}
    \centering
    \includegraphics[width=\hsize]{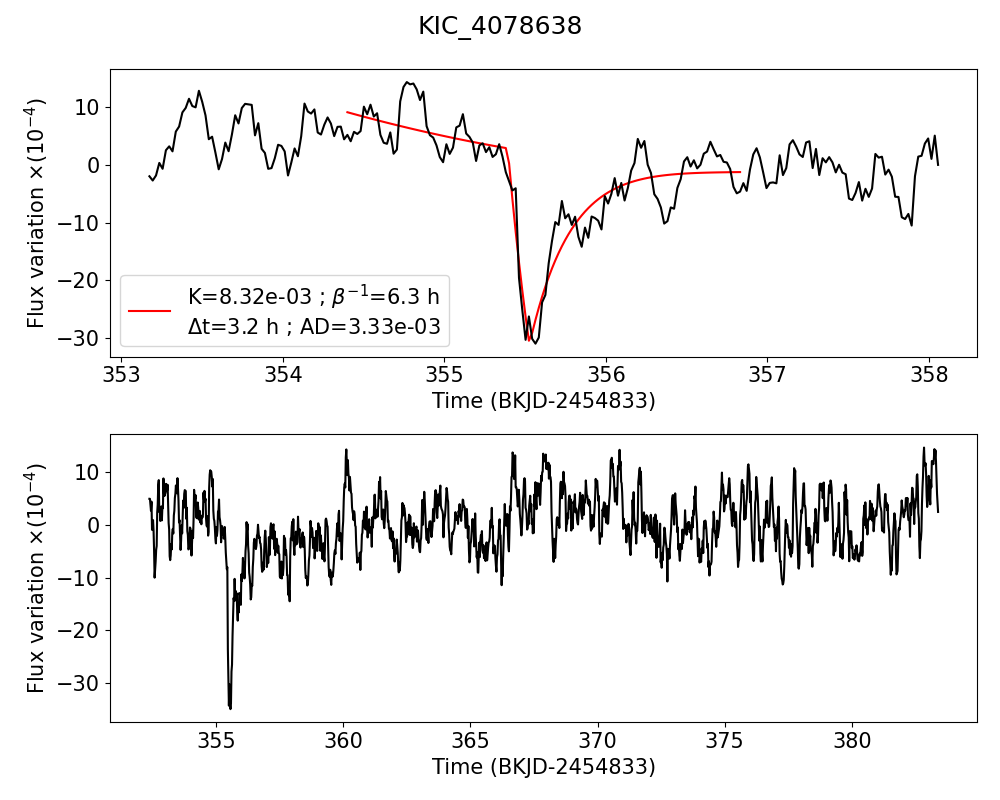}
\end{subfigure}
\begin{subfigure}{.5\hsize}
    \centering
    \includegraphics[width=\hsize]{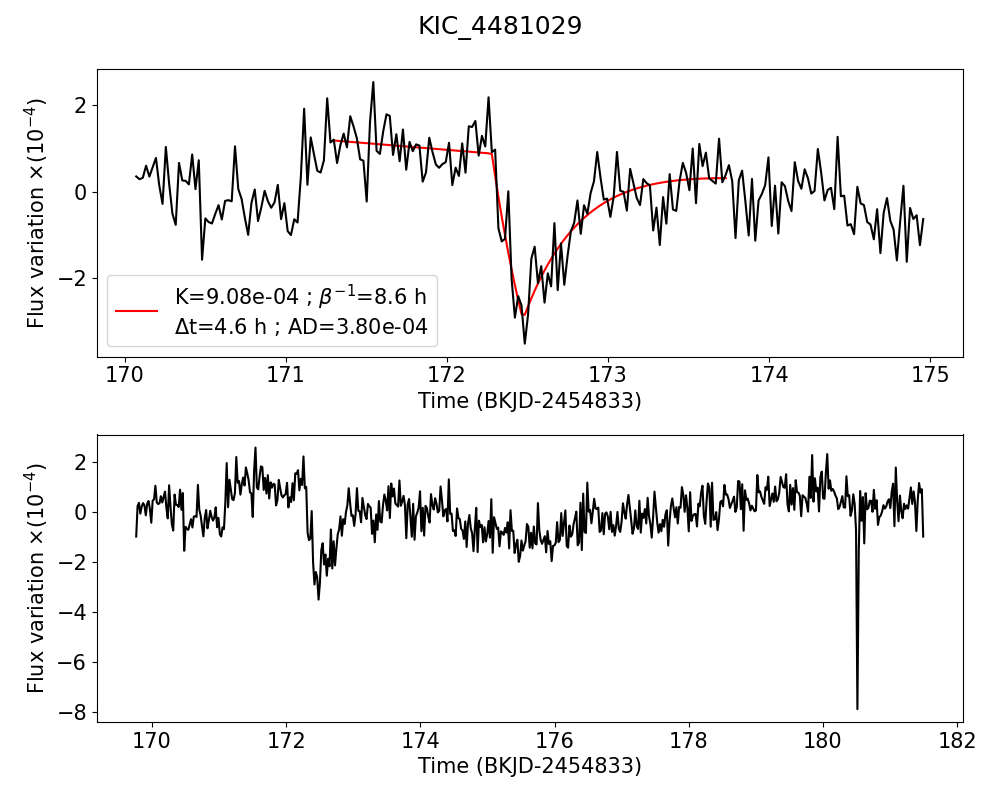}
\end{subfigure}
\begin{subfigure}{.5\hsize}
    \centering
    \includegraphics[width=\hsize]{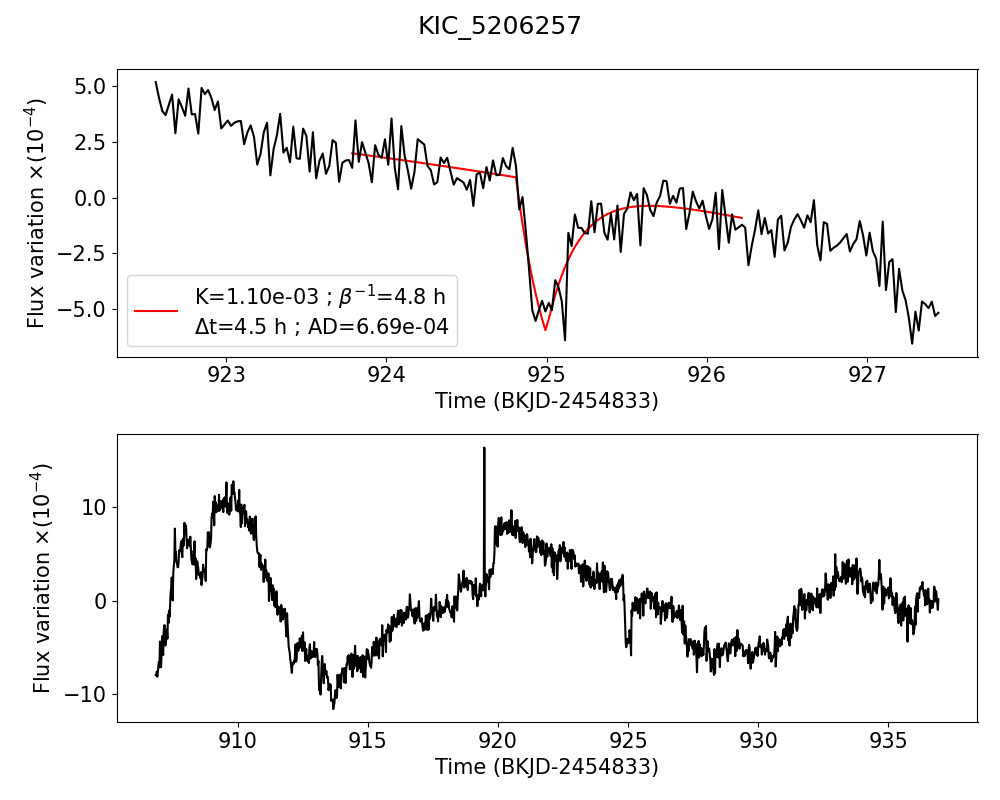}
\end{subfigure}
\begin{subfigure}{.5\hsize}
    \centering
    \includegraphics[width=\hsize]{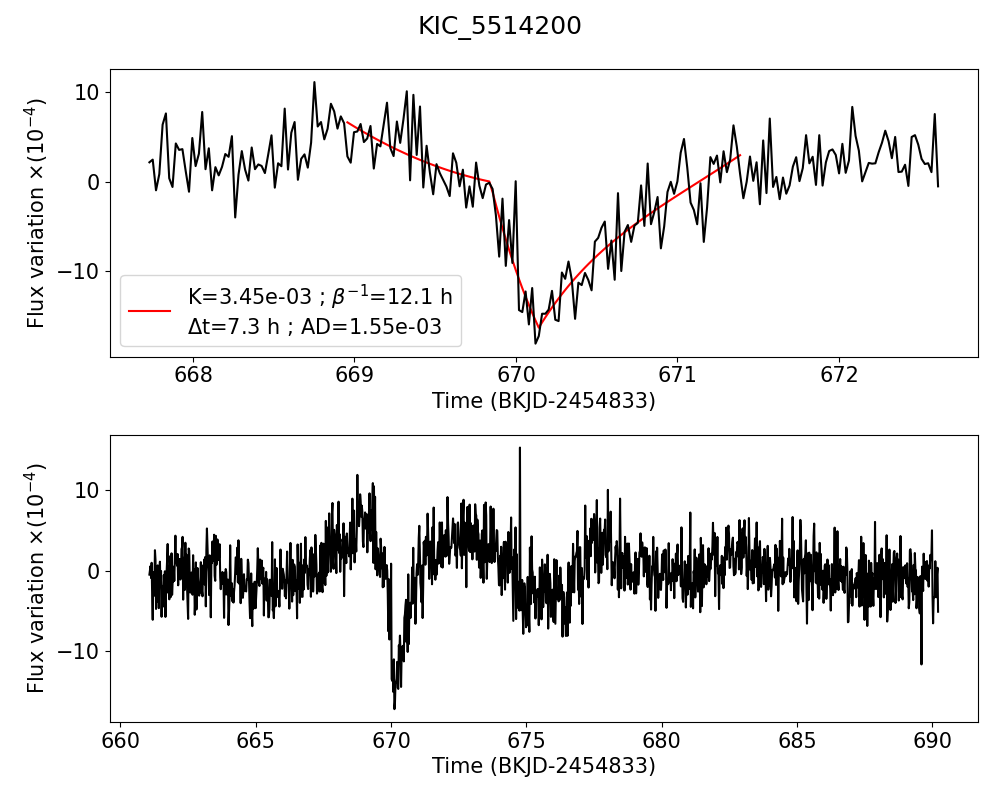}
\end{subfigure}
\begin{subfigure}{.5\hsize}
    \centering
    \includegraphics[width=\hsize]{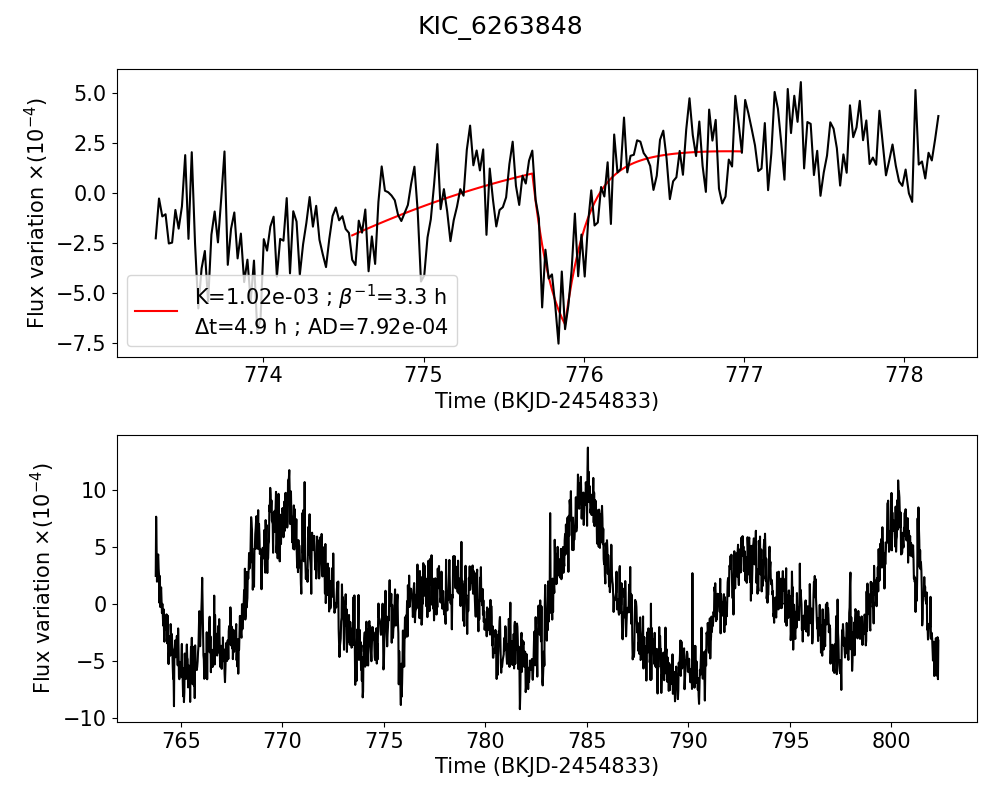}
\end{subfigure}
\begin{subfigure}{.5\hsize}
    \centering
    \includegraphics[width=\hsize]{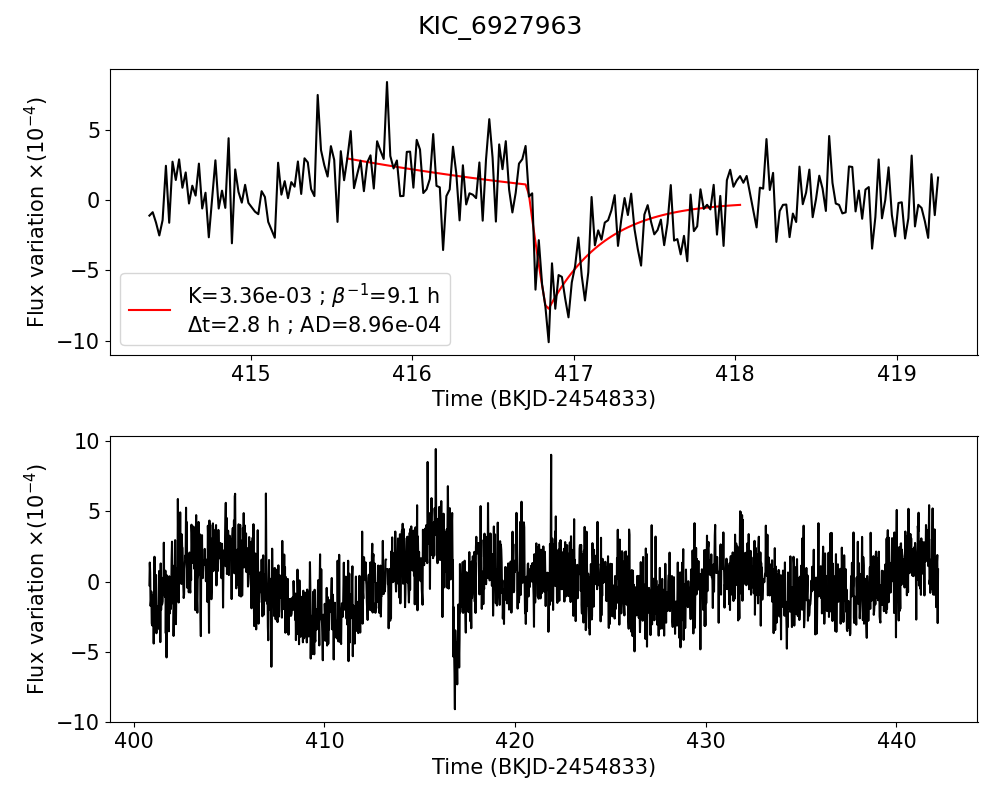}
\end{subfigure}
\end{figure*}

\begin{figure*}
\begin{subfigure}{.5\hsize}
    \centering
    \includegraphics[width=\hsize]{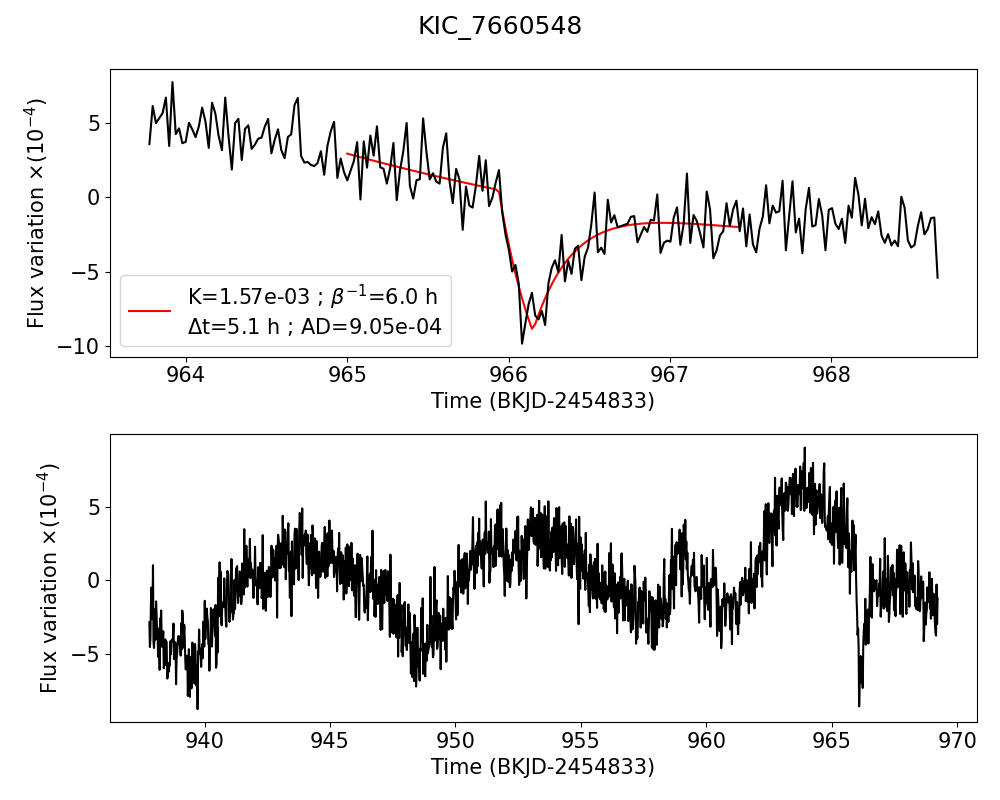}
\end{subfigure}
\begin{subfigure}{.5\hsize}
    \centering
    \includegraphics[width=\hsize]{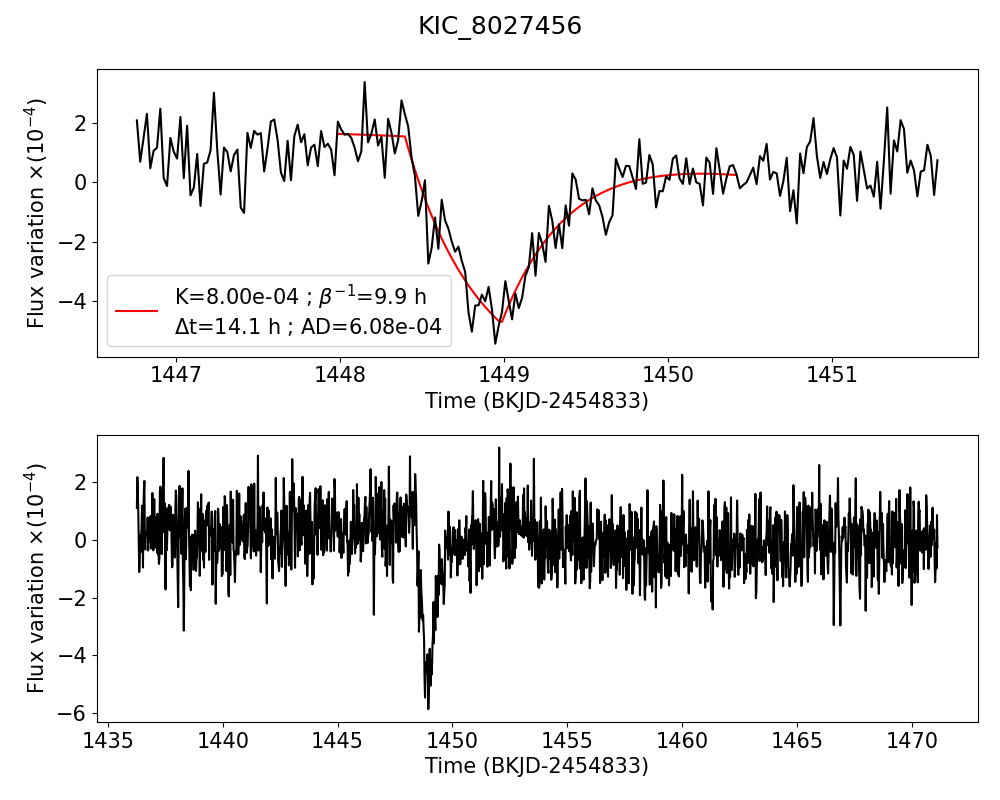}
\end{subfigure}
\begin{subfigure}{.5\hsize}
    \centering
    \includegraphics[width=\hsize]{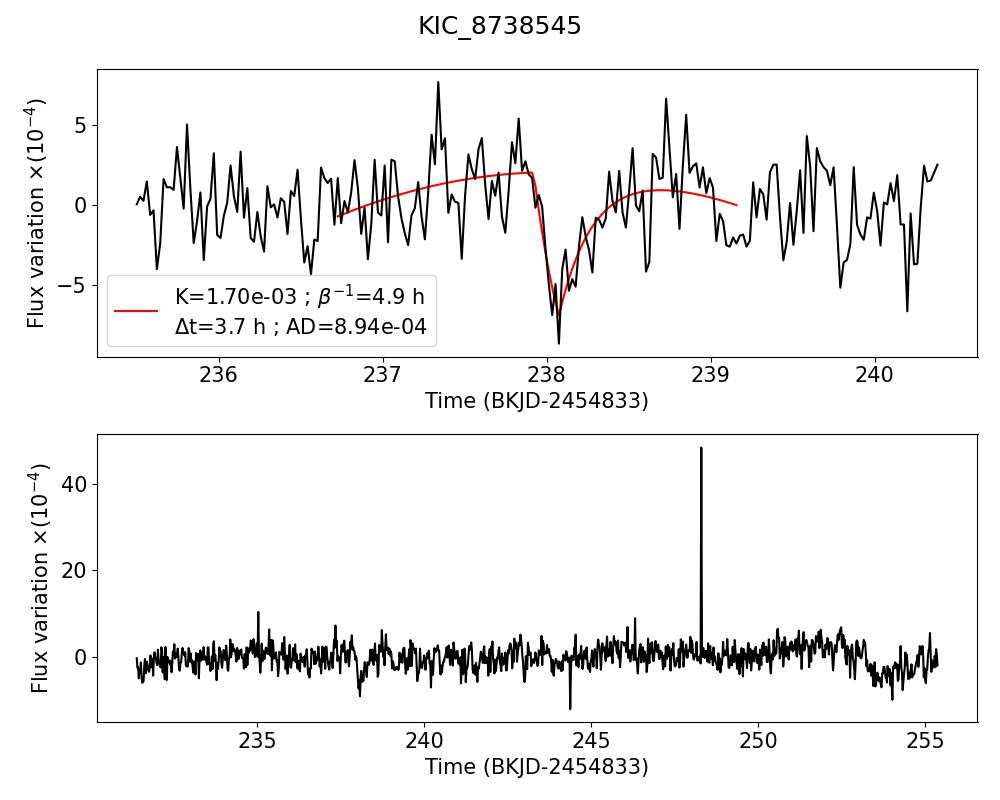}
\end{subfigure}
\begin{subfigure}{.5\hsize}
    \centering
    \includegraphics[width=\hsize]{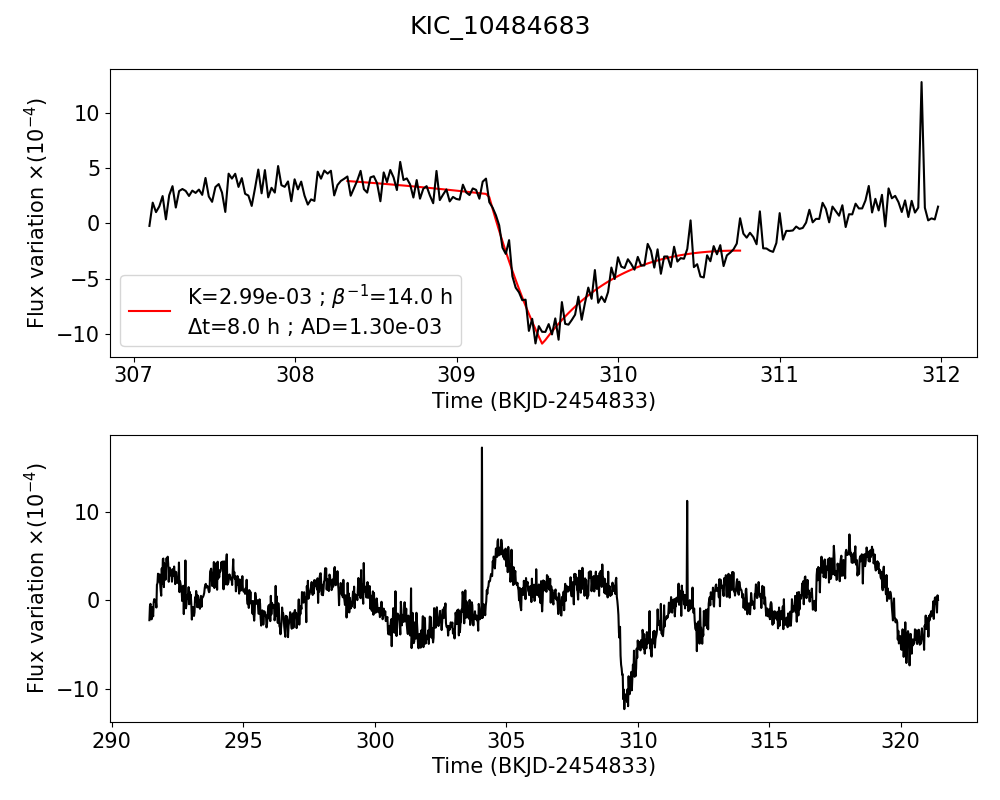}
\end{subfigure}
\begin{subfigure}{.5\hsize}
    \centering
    \includegraphics[width=\hsize]{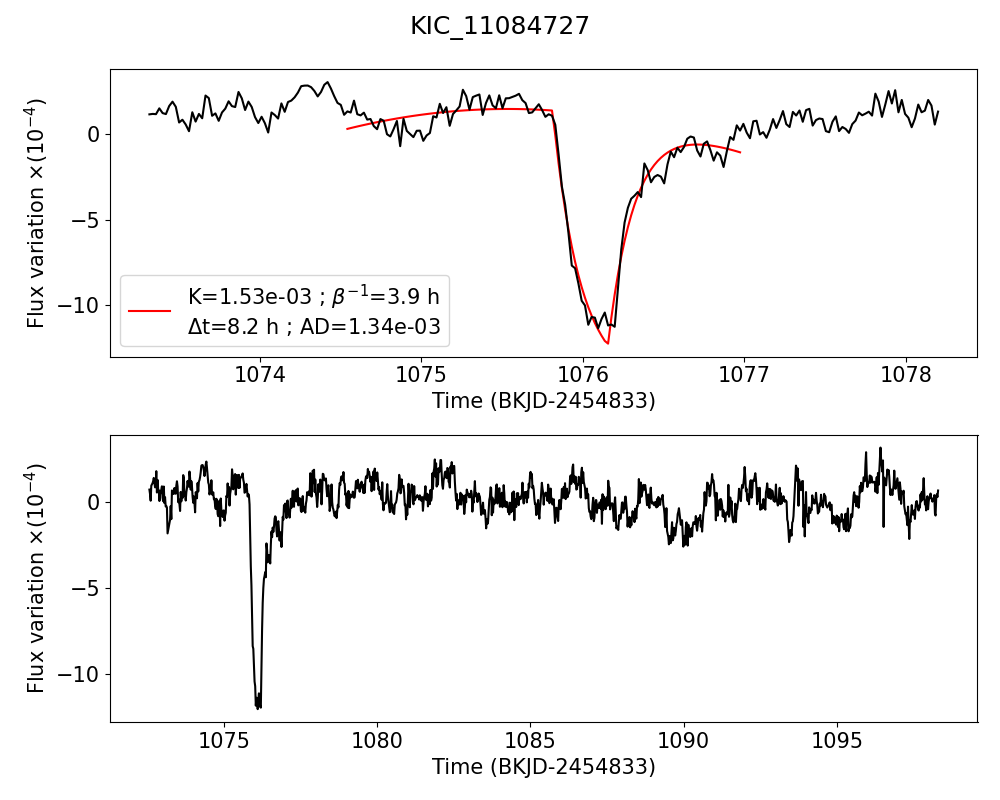}
\end{subfigure}
\end{figure*}

\FloatBarrier

\section{Possible exocometary transits light curves}
\label{Possible exocometary transits light curves}
Plots of the light curves of the possibly detected cometary transits in the second tier catalog. The plots cover a 5~days time frame centered on the transit time and with an exocomet transit fit superimposed (red line). 

\begin{figure*}[!h]

\begin{subfigure}{.5\hsize}
    \centering
    \includegraphics[width=\hsize]{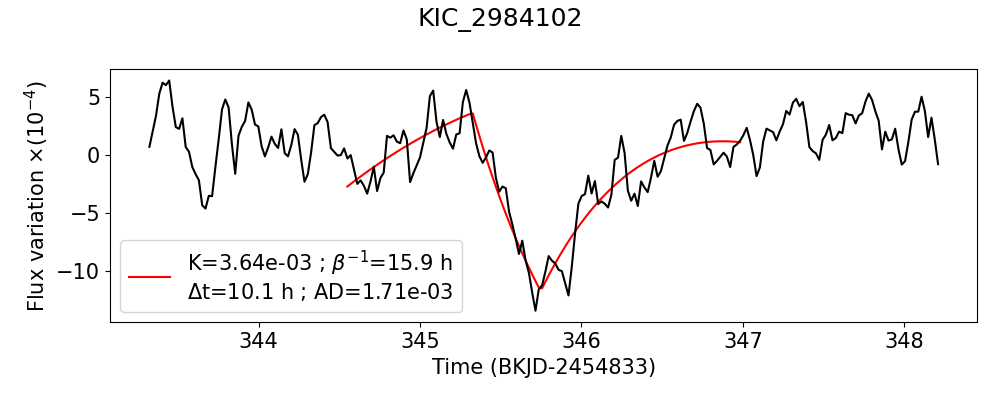}
\end{subfigure}
\begin{subfigure}{.5\hsize}
    \centering
    \includegraphics[width=\hsize]{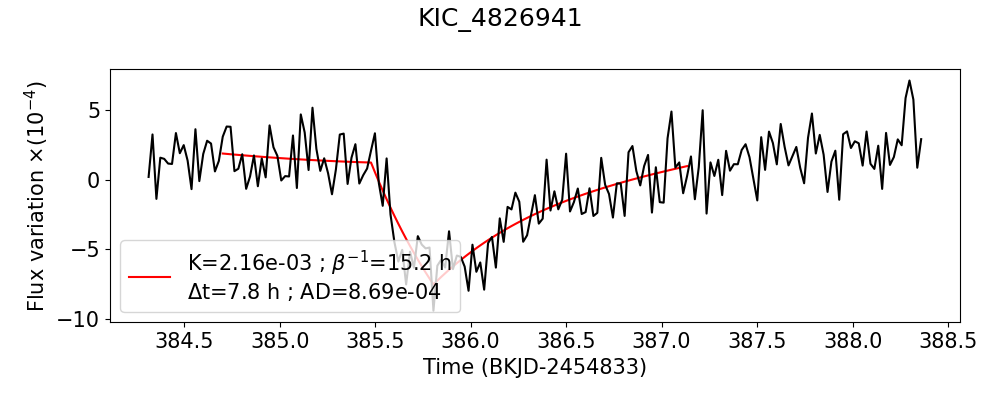}
\end{subfigure}

\begin{subfigure}{.5\hsize}
    \centering
    \includegraphics[width=\hsize]{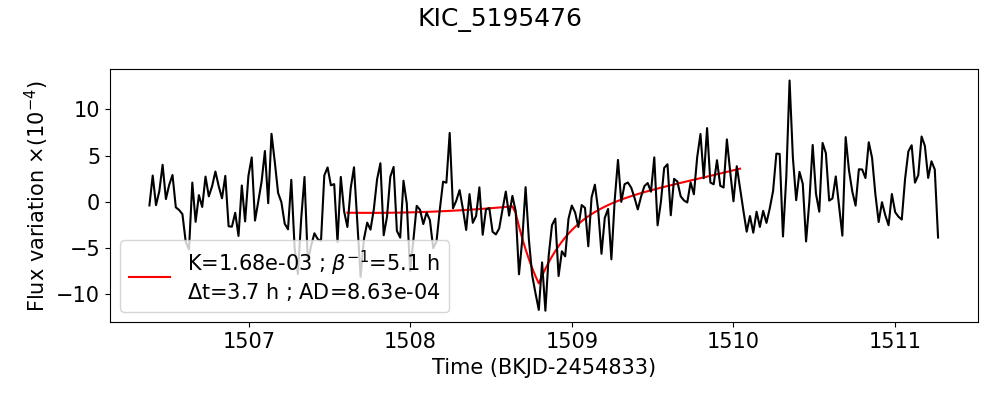}
\end{subfigure}
\begin{subfigure}{.5\hsize}
    \centering
    \includegraphics[width=\hsize]{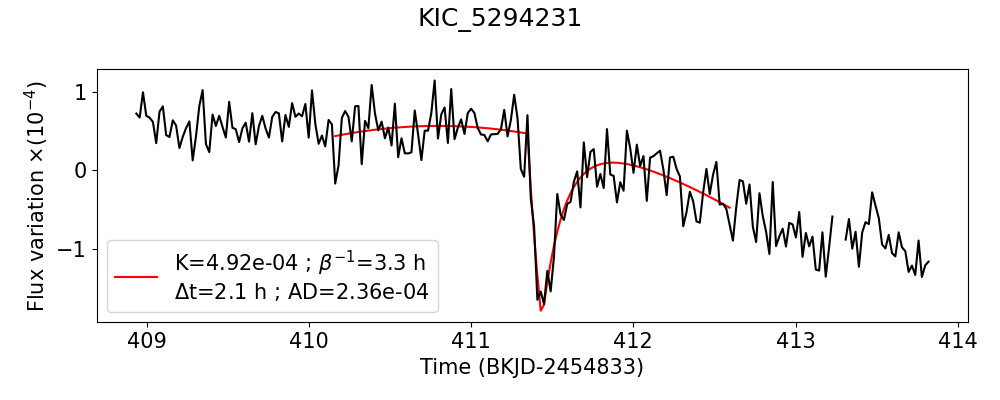}
\end{subfigure}

\begin{subfigure}{.5\hsize}
    \centering
    \includegraphics[width=\hsize]{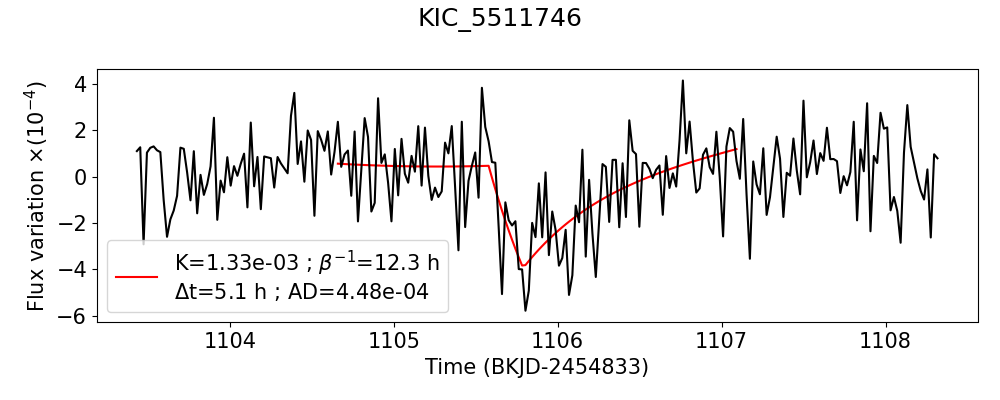}
\end{subfigure}
\begin{subfigure}{.5\hsize}
    \centering
    \includegraphics[width=\hsize]{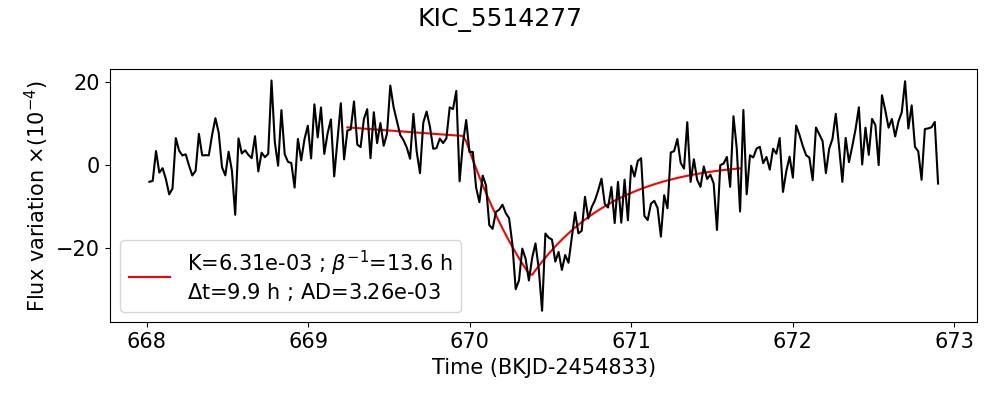}
\end{subfigure}

\begin{subfigure}{.5\hsize}
    \centering
    \includegraphics[width=\hsize]{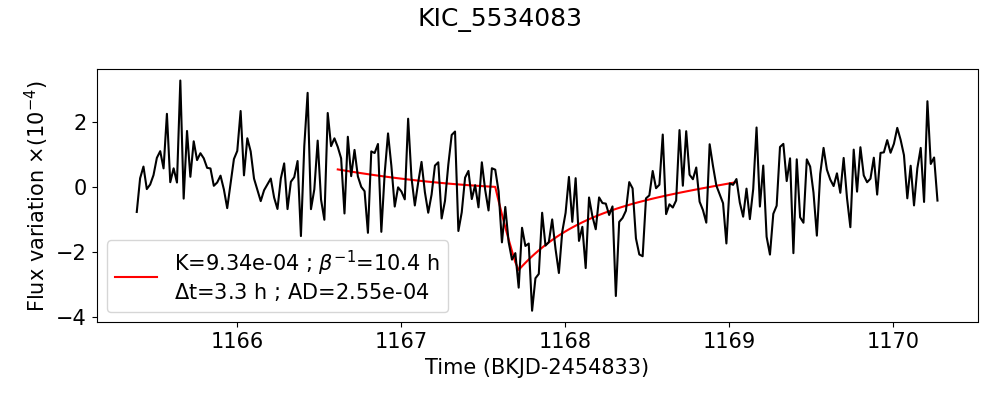}
\end{subfigure}
\begin{subfigure}{.5\hsize}
    \centering
    \includegraphics[width=\hsize]{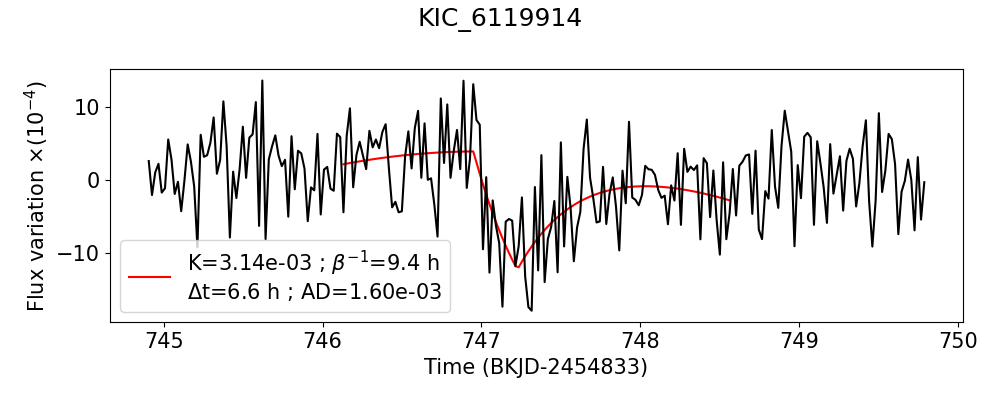}
\end{subfigure}

\begin{subfigure}{.5\hsize}
    \centering
    \includegraphics[width=\hsize]{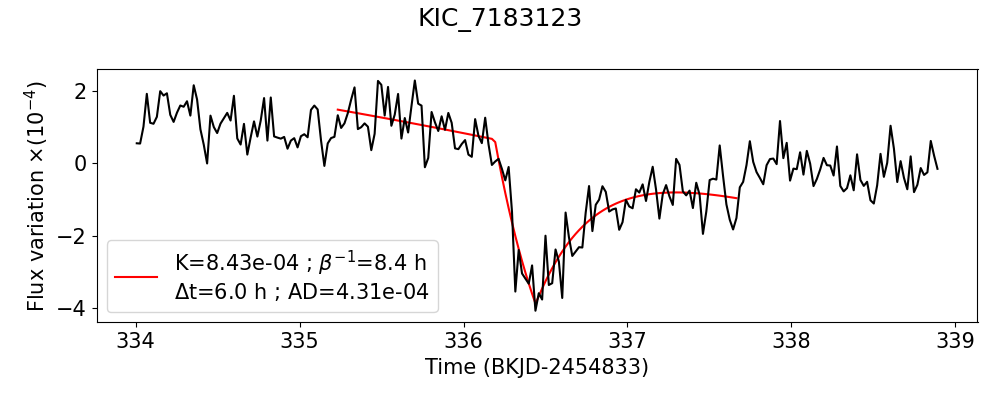}
\end{subfigure}
\begin{subfigure}{.5\hsize}
    \centering
    \includegraphics[width=\hsize]{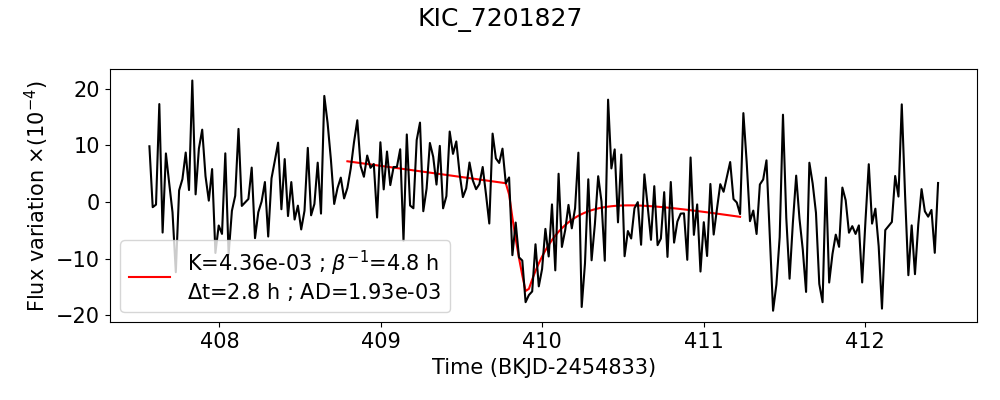}
\end{subfigure}

\begin{subfigure}{.5\hsize}
    \centering
    \includegraphics[width=\hsize]{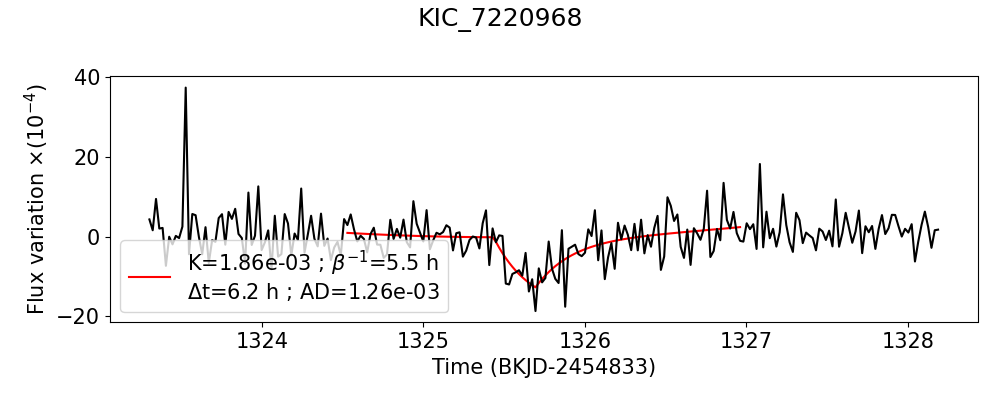}
\end{subfigure}
\begin{subfigure}{.5\hsize}
    \centering
    \includegraphics[width=\hsize]{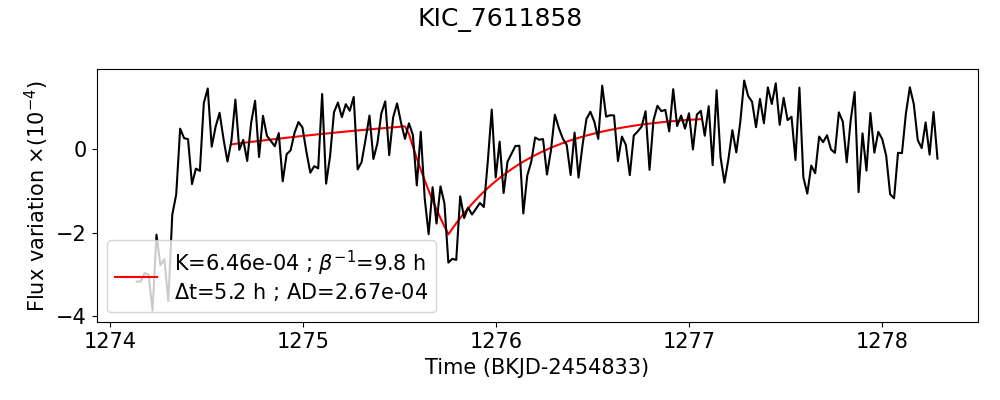}
\end{subfigure}

\end{figure*}

\begin{figure*}[!h]

\begin{subfigure}{.5\hsize}
    \centering
    \includegraphics[width=\hsize]{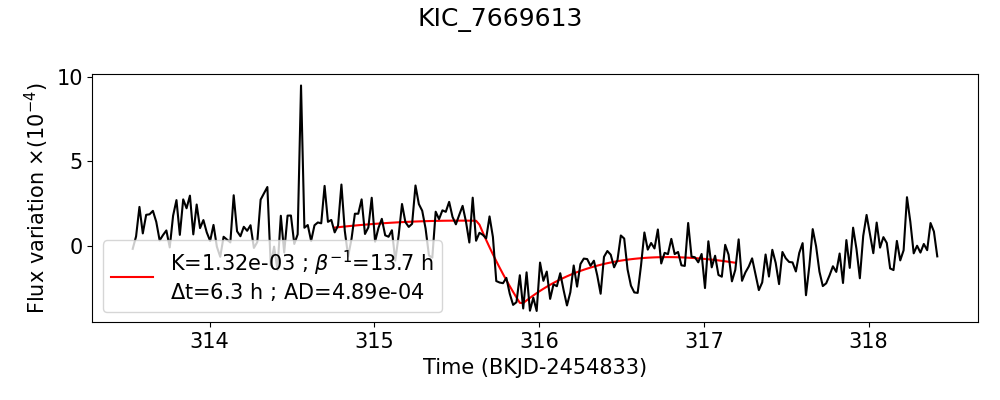}
\end{subfigure}
\begin{subfigure}{.5\hsize}
    \centering
    \includegraphics[width=\hsize]{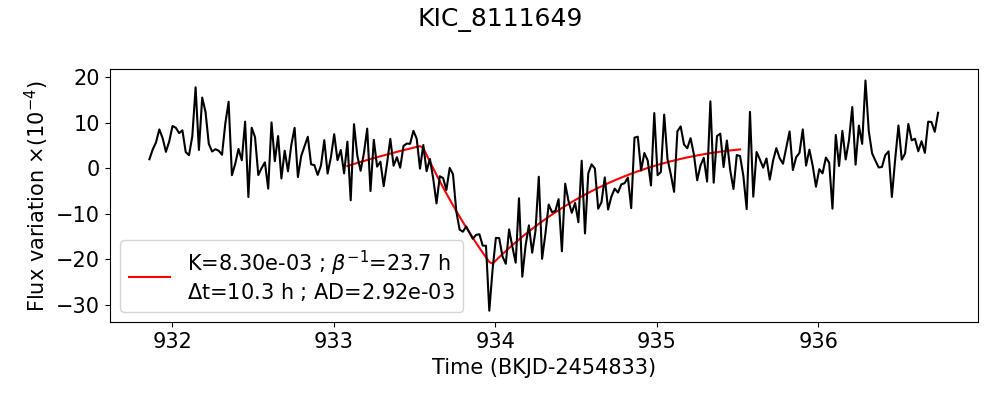}
\end{subfigure}

\begin{subfigure}{.5\hsize}
    \centering
    \includegraphics[width=\hsize]{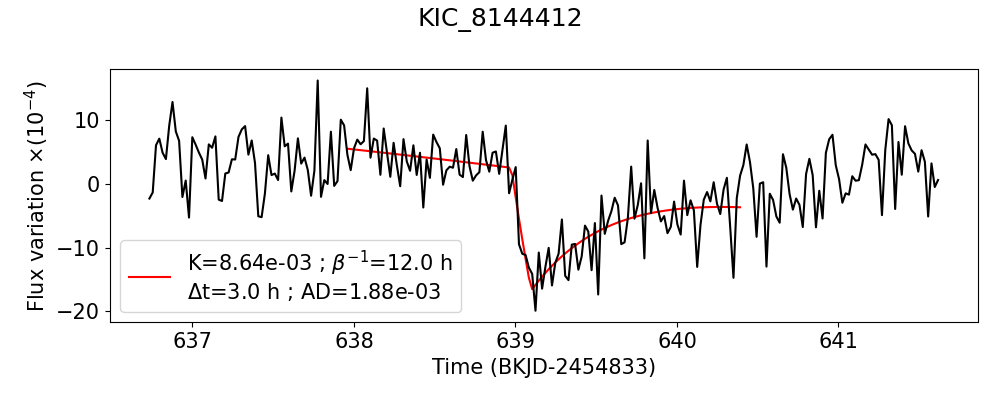}
\end{subfigure}
\begin{subfigure}{.5\hsize}
    \centering
    \includegraphics[width=\hsize]{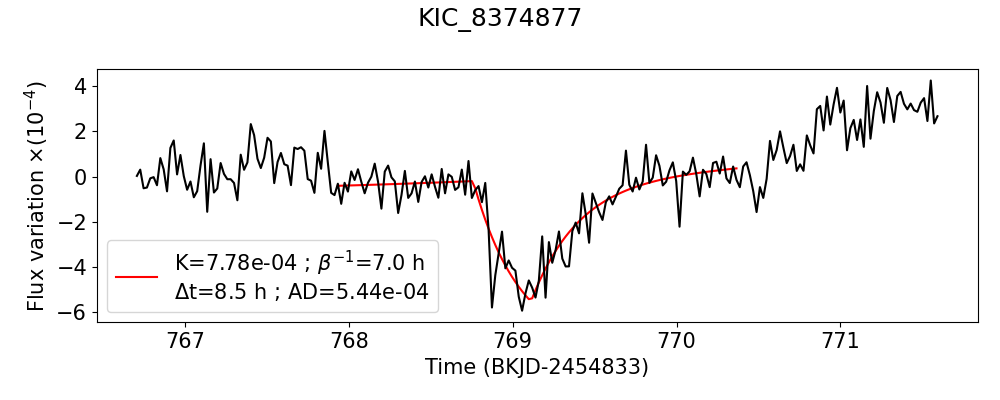}
\end{subfigure}

\begin{subfigure}{.5\hsize}
    \centering
    \includegraphics[width=\hsize]{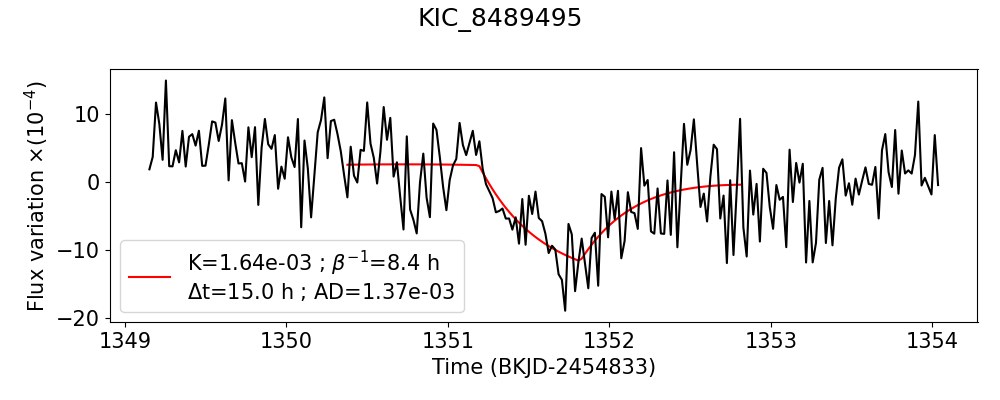}
\end{subfigure}
\begin{subfigure}{.5\hsize}
    \centering
    \includegraphics[width=\hsize]{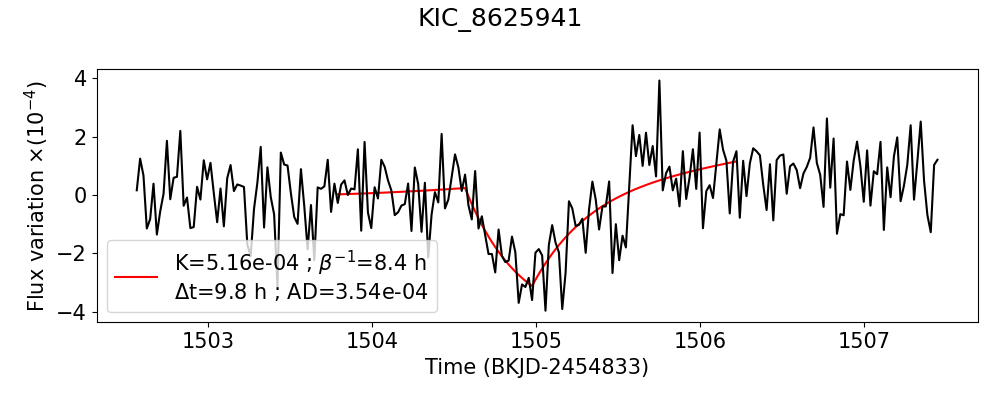}
\end{subfigure}

\begin{subfigure}{.5\hsize}
    \centering
    \includegraphics[width=\hsize]{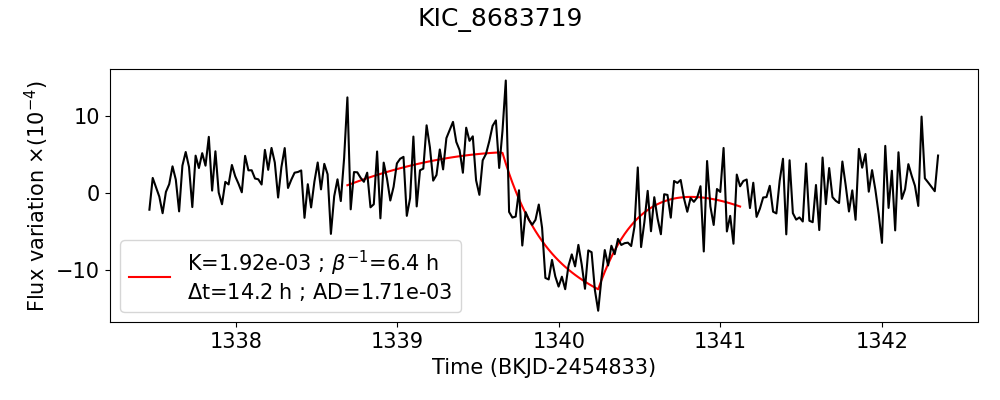}
\end{subfigure}
\begin{subfigure}{.5\hsize}
    \centering
    \includegraphics[width=\hsize]{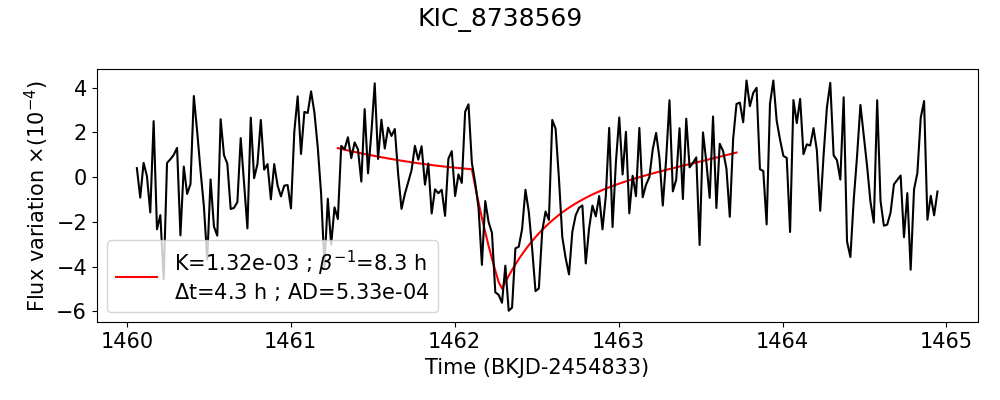}
\end{subfigure}

\begin{subfigure}{.5\hsize}
    \centering
    \includegraphics[width=\hsize]{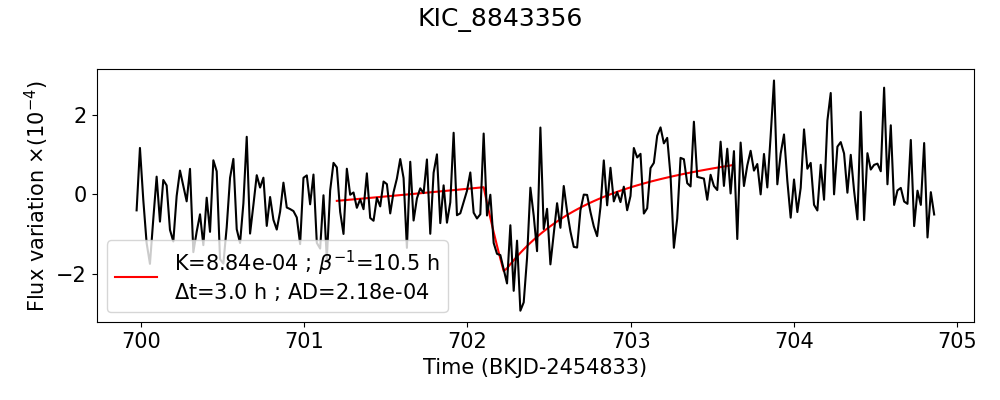}
\end{subfigure}
\begin{subfigure}{.5\hsize}
    \centering
    \includegraphics[width=\hsize]{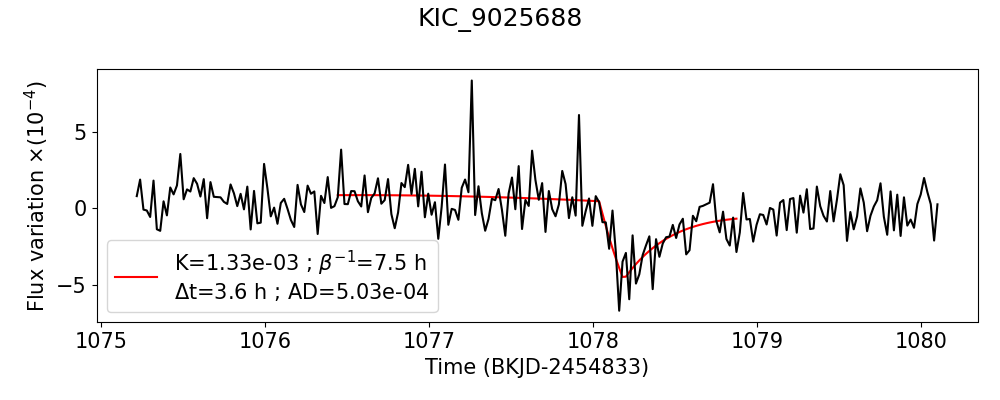}
\end{subfigure}

\begin{subfigure}{.5\hsize}
    \centering
    \includegraphics[width=\hsize]{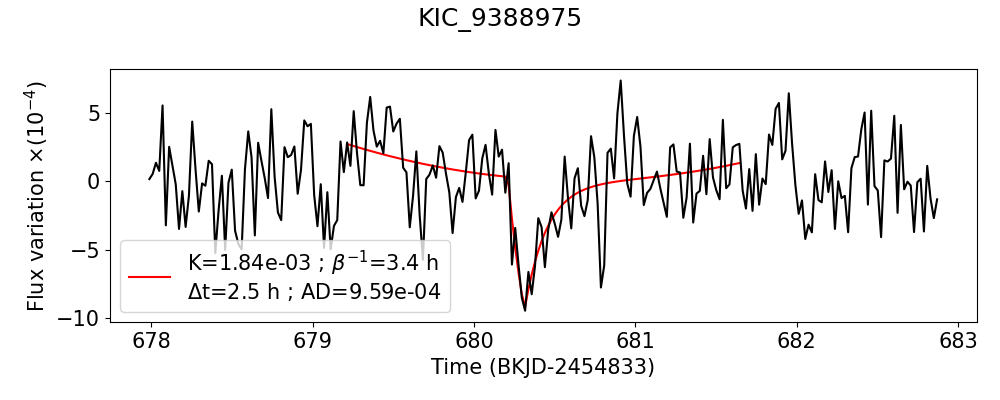}
\end{subfigure}
\begin{subfigure}{.5\hsize}
    \centering
    \includegraphics[width=\hsize]{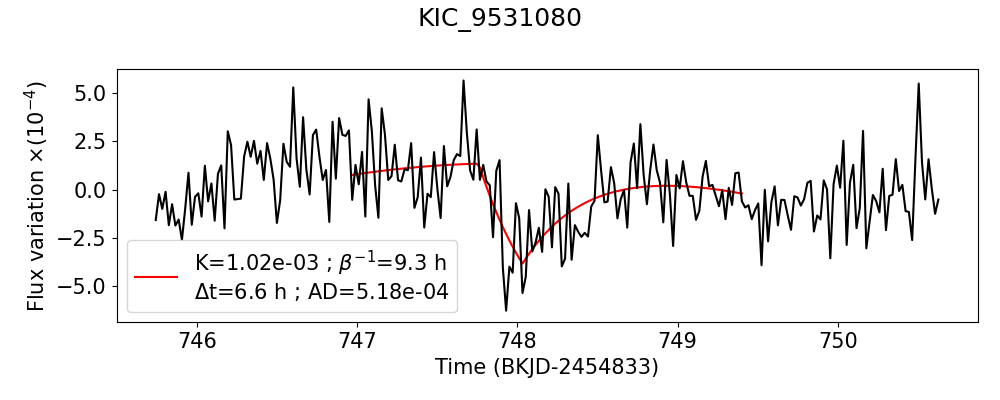}
\end{subfigure}

\end{figure*}

\begin{figure*}[!h]

\begin{subfigure}{.5\hsize}
    \centering
    \includegraphics[width=\hsize]{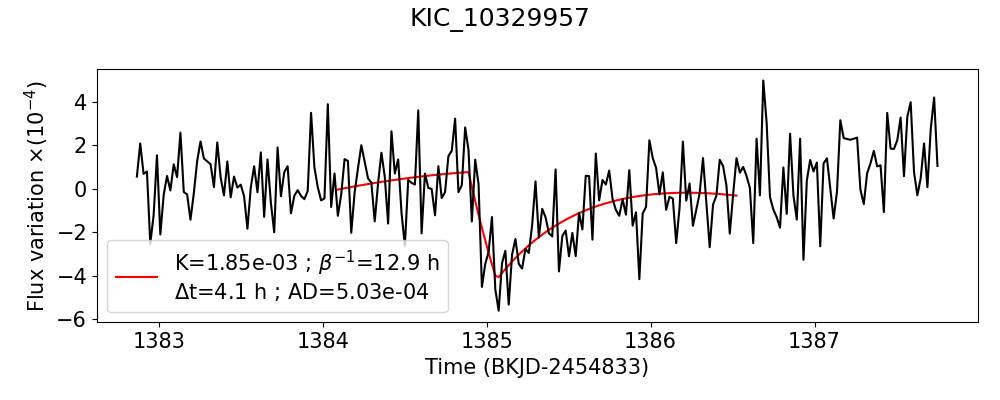}
\end{subfigure}
\begin{subfigure}{.5\hsize}
    \centering
    \includegraphics[width=\hsize]{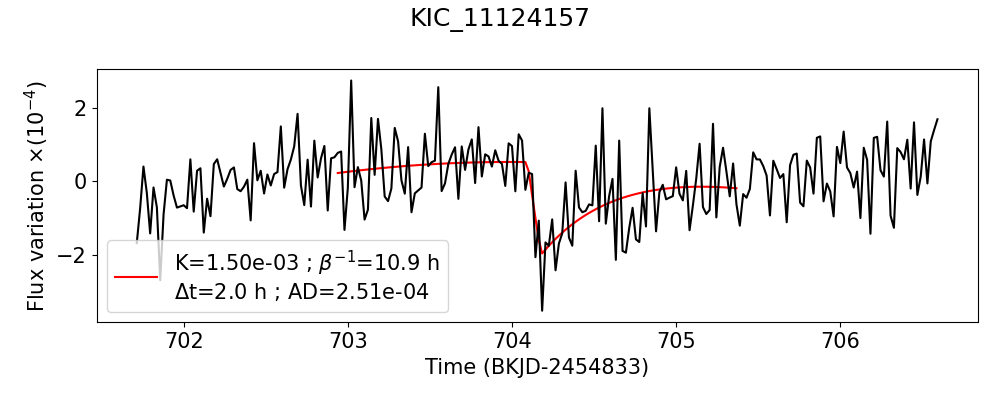}
\end{subfigure}

\begin{subfigure}{.5\hsize}
    \centering
    \includegraphics[width=\hsize]{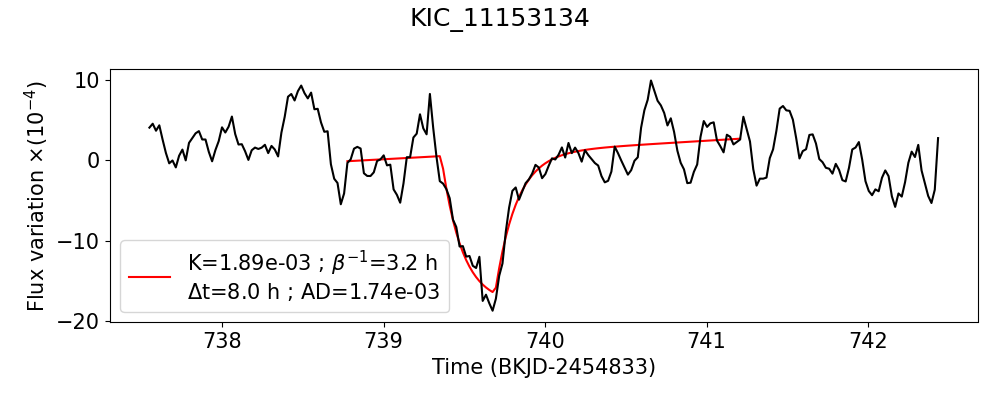}
\end{subfigure}
\begin{subfigure}{.5\hsize}
    \centering
    \includegraphics[width=\hsize]{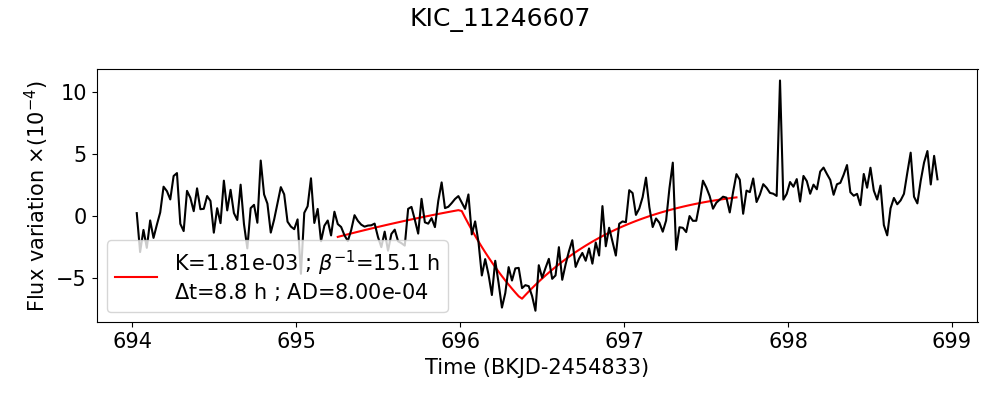}
\end{subfigure}

\begin{subfigure}{.5\hsize}
    \centering
    \includegraphics[width=\hsize]{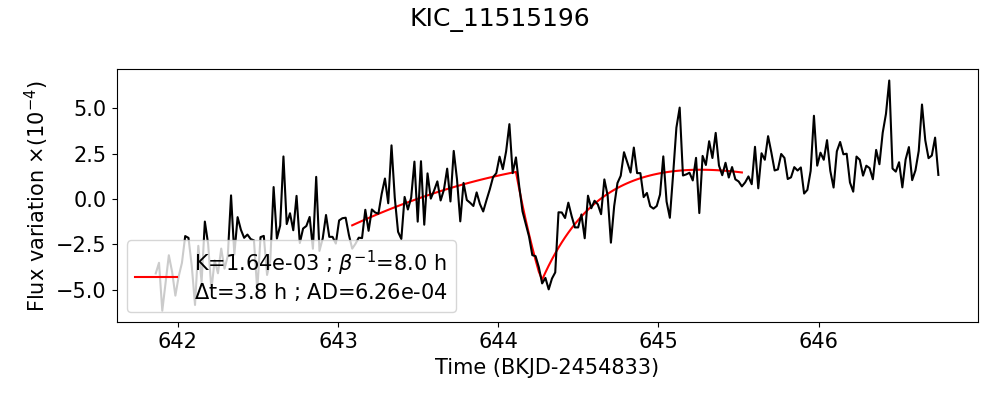}
\end{subfigure}
\begin{subfigure}{.5\hsize}
    \centering
    \includegraphics[width=\hsize]{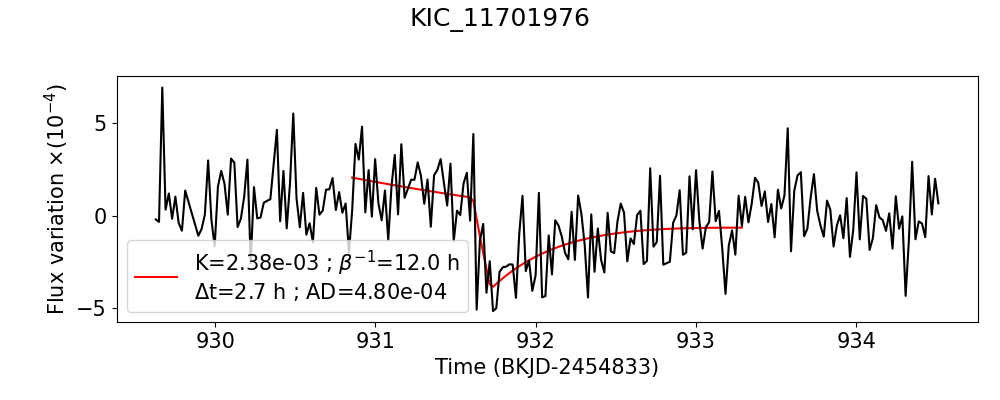}
\end{subfigure}

\end{figure*}
   
\end{appendix}

\end{document}